\documentclass{article}[12pt,a4paper]
\usepackage{graphicx}  
\usepackage{dcolumn}   
\usepackage{bm}        
\usepackage{amssymb}   
\usepackage{amssymb,epsfig,subfigure}
\usepackage[]{amsmath,amssymb}
\usepackage{graphics,epsfig}
\usepackage[latin1]{inputenc}
\usepackage[T1]{fontenc}
\usepackage{amsfonts}
\usepackage{latexsym}
\usepackage[english]{babel}

\usepackage{cite}
\usepackage{color}
\usepackage{geometry}

\usepackage{hyperref}
\usepackage{comment}
\usepackage[font=footnotesize]{caption}

\usepackage{esint}
\usepackage{cancel}
\usepackage{verbatim}
\newcommand{\be}{\begin{equation}}
\newcommand{\ee}{\end{equation}}
\newcommand{\beq}{\begin{equation}}
\newcommand{\eeq}{\end{equation}}
\newcommand{\bea}{\begin{eqnarray}}
\newcommand{\eea}{\end{eqnarray}}

\def\be{\begin{equation}}
\def\ee{\end{equation}}
\def\ba{\begin{eqnarray}}
\def\ea{\end{eqnarray}}

\numberwithin{equation}{section}

\begin{document}

\date{\empty}

\title{Entropy and modular Hamiltonian for a free chiral scalar in two intervals}

\author{Ra\'{u}l E. Arias$^{a}$, Horacio Casini$^b$, Marina Huerta$^b$, Diego Pontello$^b$}

\maketitle

\begin{center}

{\sl $^a$ Instituto de F\'{\i}sica de La Plata - CONICET\\
C.C. 67, 1900 La Plata, Argentina.}

~

{\sl $^b$ Centro At\'omico Bariloche,\\
8400-S.C. de Bariloche, R\'{\i}o Negro, Argentina}

\end{center}

\maketitle

\begin{abstract}
We calculate the analytic form of the vacuum modular Hamiltonian for a two interval region and the algebra of a current $j(x)=\partial \phi(x)$ corresponding to a chiral free scalar $\phi$ in $d=2$. We also compute explicitly the mutual information between the intervals. This model shows a failure of Haag duality for two intervals that translates into a loss of a symmetry property for the mutual information usually associated with modular invariance. Contrary to the case of a free massless fermion, the modular Hamiltonian turns out to be completely non local.  
The calculation is done diagonalizing the density matrix by computing the eigensystem of a correlator kernel operator. These eigenvectors are obtained by a novel method that involves solving an equivalent problem for an holomorphic function in the complex plane where multiplicative boundary conditions are imposed on the intervals. Using the same technique we also re-derive the free fermion modular Hamiltonian in a more transparent way.  
\end{abstract}

\section{Introduction}
The reduced state to a local algebra of operators in a region in quantum field theory (QFT) can be expressed, in presence of an ultraviolet cutoff, as a density matrix $\rho=e^{-{\cal H}}$. The exponent ${\cal H}$, called the modular Hamiltonian, conveniently encodes the reduced state. It retains a meaning in the continuum limit as the generator of unitary transformations corresponding to imaginary powers of the density matrix, $\rho^{i \tau}=e^{-i H \tau}$, the so called modular flow. 

The structural importance of modular flows in the algebraic formulation of QFT has been being recognized since early times \cite{haag}. More recently, statistical properties of reduced states in QFT, have been the subject of much interest. In particular, entropy and relative entropy have simple geometric duals for holographic QFT \cite{duals1,duals2}. In this context, the modular Hamiltonian in the boundary theory and its bulk dual have been used to elucidate localization properties of degrees of freedom in quantum gravity \cite{moduholo1,moduholo2,moduholo3}. More generaly, knowledge of the modular Hamiltonian is an important step in relative entropy calculations, and it is fundamental to the formulation of entropy bounds \cite{bounds1,bounds2,bounds3,bounds4,bounds5} and the proof of several energy conditions \cite{energy1,energy2,energy3,energy4,energy5}.  

However, most of our knowledge of the explicit form of modular Hamiltonians reduce to some examples where the modular flow is local, and it is primarily determined by spacetime symmetries \cite{spacetime1,spacetime2,spacetime3}. 
 In more generality, it is possible to identify a local part of the modular Hamiltonian which should have a large degree of universality \cite{ante,loctemp}, while not much is known about non local terms. An example of non local modular Hamiltonian which has been explicitly computed is for the vacuum state  of the free massless fermion in $d=2$ \cite{modu} (see also \cite{rehren, wong}). In this case ${\cal H}$ for several disjoint intervals has a local term proportional to the energy density and an additional non local part given by a quadratic expression in the fermion field. This last term however, does not contain  all possible products of pairs $\psi^\dagger(x) \psi(y)$ for fields located in arbitrary points $x,y$ in the region, but only selected points appear: For each $x$ in one interval only one specific ``conjugate'' point $y$ appears in any of the other intervals.    

There has been much progress in understanding local statistical properties of the vacuum reduced to two intervals in more general CFT. The Renyi entropies for integer index $n$ have been explicitly computed for several models of interest \cite{renyis1,renyis2,renyis3,yapeyu1,yapeyu2,yapeyu3}. However, analytic continuation to $n\rightarrow 1$ to obtain the entanglement entropy has shown to be a difficult task. The computation of modular Hamiltonians of the vacuum for two intervals has also proved elusive \cite{elusive}.      

A natural candidate to apply kernel methods is the free massless scalar. However, in two dimensions the uncompactified scalar field itself is ill defined due to infrared divergences, and one has to restrict the algebra to the derivatives of the scalar field. In this paper we compute the entanglement entropy and the modular Hamiltonian explicitly for two intervals in the theory of a chiral current, that is, the chiral derivative of the scalar field. The model is free and the modular Hamiltonian is quadratic. We diagonalize the kernel in this quadratic expression. The eigenvectors are obtained by an adaptation of the method in \cite{ante}. In the present case the problem is maped to one of finding analytic functions in the complex plane with specific multiplicative boundary conditions on the two intervals, located on the real line. In contrast to the case of the fermion, for the chiral scalar, the modular Hamiltonian is completely non local, while it still contains the expected local term proportional to the energy density operator \cite{loctemp}.  

The mutual information turns out to be a function of the cross ratio of the four end points of the intervals given by an integral over Hypergeometric functions. We check the result with numerical simulations on a lattice. For global pure states it is naturally expected that the entanglement entropy for complementary regions coincide. In this case the mutual information for two intervals would acquire an additional symmetry property relating the cross ratios $\eta\in (0,1)$ and $(1-\eta)$ \cite{cteo}. This property also follows from modular invariance in the replica trick calculation with Euclidean path integrals \cite{moduinvar}. In the present model this symmetry is absent. This is explained by failure of Haag duality for two intervals: The algebra corresponding to the complement of the two intervals is smaller than the commutant of the algebra of the two intervals. Kawahigashi, Longo and M\"uger related this failure of Haag duality for two intervals in chiral conformal models to an algebraic index ($\mu$-index) on inclusion of subalgebras \cite{kawa}. This index also determines the amount of asymmetry in the mutual information \cite{longo}. For the chiral scalar  the $\mu$-index is infinity. 

Our new method to deal with eigenvectors of the correlator kernels also leads to a better understanding of the derivation of the free fermion modular Hamiltonian. We start by a detailed derivation of this result in the next section. This will also allow us to introduce the main ideas to be used in the more complicated case of the chiral scalar. However, the treatment of the scalar case is self contained and a reader not interested in the discussion of the fermion field can start directly with section \ref{escalar}.  
 
In section \ref{escalar} we describe the algebra of the chiral scalar current, and the relevant kernels. In section \ref{unintervalo} we show for a one interval region the kernel diagonalization, and the calculation of the entropy and modular Hamiltonian. The case of two intervals is dealt with in section \ref{dosintervalos}, where we also compute the mutual information and modular Hamiltonian. Section \ref{failure} discuss the reasons for the breaking of the symmetry property of the mutual information. In section \ref{numerico} we present the numerical calculations in a lattice and compare with the analytic results for the mutual information.  Finally we end with a brief summary in section \ref{conclusions}.

\section{The massless free fermion revisited}
A complete description of the reduced density matrix of a massless fermion field for multi-interval regions was given in \cite{modu}. This was achieved by diagonalyzing the correlator kernel in the region, using previous results on the literature about singular kernels of the Cauchy type \cite{ru}. Here we are doing this diagonalization transparent by mapping the problem of integral equations in one dimension to one of partial differential equations in two dimensions, following \cite{ante}. The form of the eigenvectors as well as its main properties are easily derived using this trick. This will also show how to generalize this calculation to scalars. We will treat the scalar field in the next section.

Since chiralities decouple in the massless case we consider a chiral fermion in $d=2$, which only depends on a null coordinate. In our notation the variables $x$, $y$, etc, correspond to this null coordinate. We will consider a region $A=(a_1,b_1)\cup (a_2,b_2)\cup\hdots \cup (a_n,b_n)$ formed by $n$ intervals. The field satisfies the anticommutation relations $\{\psi(x),\psi^\dagger(y)\}=\delta(x-y)$. The correlator kernel is\footnote{Changing the sign of the imaginary part in this expression corresponds to changing chirality.}
\be
C(x-y)=\langle 0|\psi(x)\psi^\dagger(y)|0\rangle= \frac{1}{2}\delta(x-y)+\frac{i}{2\pi} \frac{1}{x-y} \,, \label{corrC}
\ee
where the distribution on the left hand side is understood in principal value regularization.  
This is a projector when acting on the full line, and on the region $A$ is an Hermitian operator with continuous eigenvalues in the range $(0,1)$ \cite{modu}. To obtain the modular Hamiltonian it is important to solve the spectrum of the correlator reduced to the region $C_A(x,y)\equiv C(x,y)|_{x,y\in A}$, because the modular Hamiltonian is given by \cite{review,peschel,araki}
\be
{\cal H}=\int_A dx\,dy\, \psi^\dagger(x) H(x,y) \psi(y)\,,\label{citric}
\ee
where the kernel $H$ is 
\be
H=-\log(C_A^{-1}-1)\,.\label{hache}
\ee
This last equation is understood as an operator equation, where the action of the operators is defined through their kernels.

\subsection{An equivalent problem in the complex plane} \label{cp}

In this section we will relate the orginal problem of solving the spectrum of $C_A$ as a kernel in $A$, to a new problem about a function in the complex plane. At the end, we will arrive at the same results of \cite{ante}, but here we adapt the discussion to the chiral fermion field. 

For such purpose,  we think the $n$ intervals $A$ as included in the real axis of the complex plane.  For each $\lambda \in \mathbb{R}$ consider the following problem about a function $S(z)$ in the complex plane
\bea
&& S(z) \text{ analytic in }\mathbb{C}-\bar{A} \, , \label{hol_A} \\
&& S^{+}(x_1)=\lim_{x_2\rightarrow 0^+} S(x_1+ i x_2)=\lambda \lim_{x_2\rightarrow 0^-} S(x_1+ i x_2)=\lambda\, S^-(x_1) \, , \hspace{1cm} x_1\in A, \label{disc} \\
&& \lim_{z\rightarrow \infty} \left| z \, S(z) \right| <\infty\,, \label{bc1} \\
&& \lim_{z \rightarrow \partial A} l_{z,\partial A} \,S(z)\rightarrow 0 \, ,\label{bc}
\eea
 where $l_{z,\partial A}$ is the distance from $z$ to the boundary $\partial A$ (formed by $2n$ disjoint points). Thus, $S(z)$ has a cut over $A$ with multiplicative boundary conditions. Consider now the complex integral
\be
\oint dz_2 \, \frac{1}{z_2-z_1} S(z_2)\, . \label{circ_int}
\ee
where we choose an integration contour that encircles both $A$ and $z_1$ in the positive (anticlockwise) direction. Then the integral vanishes because of (\ref{bc1}), but writing it as two separated contributions from the pole at $z_1$ and the integration around the cut $A$ we get  
\be
S(z_1)=\frac{1}{ 2 \pi i} \int_A dy\,  \frac{1}{y-z_1} (S^+(y)-S^-(y))=\frac{1-\lambda^{-1}}{ 2 \pi i} \int_A dy\,  \frac{1}{y-z_1} S^+(y)\, ,\label{todosla}
\ee
where we have used the boundary condition (\ref{disc}). We remark there are no contributions from the end points of the intervals due to (\ref{bc}).
This equation gives the value for $S(z)$ on any point $z \in \mathbb{C}$ from its values at the cut $A$.  Taking the limit $z_1 \rightarrow x \in A$ from above, and using 
\be
\lim_{y\rightarrow 0^+} \frac{1}{x+ i y}= \frac{1}{x}-i \pi \delta(x)\,, \label{plem}
\ee
we get
\be
\int_A dy\, C_A(x-y) S^+(y)=\frac{\lambda}{\lambda-1} S^+(x)\, ,\label{coco}
\ee
which means that the boundary value of $S(z)$ plays the role of an eigenvector with eigenvalue $\lambda(\lambda-1)^{-1}$ for the correlator kernel on $A$. 
Since the spectrum of $C_A(x-y)$ is restricted to $(0,1)$ (see \cite{review}), we have that $\lambda \in (-\infty,0)$. For later convenience we write
\be
\lambda = - \mathrm{e}^{2 \pi s}\, , \hspace{0.5 cm}  s \in \mathbb{R} \, . \label{lami}
\ee

Conversely, suppose we have a solution $S^+(x)$ of the eq. (\ref{coco}) for some $\lambda \in \mathbb{R}^-$ with appropriate boundary conditions on the end points of the intervals as in (\ref{bc}).\footnote{This is precisely the boundary condition of the eigenvectors for the vacuum state \cite{review}.} Then equation (\ref{todosla}) gives a complex valued function $S(z)$ satisfying all the properties (\ref{hol_A}-\ref{bc}). For the boundary condition (\ref{disc}), the function $S(z)$ defined in this way has boundary value $S^+(x)$ at the upper side of the cut, and for the lower side of the cut we have to use   
\be
\lim_{y\rightarrow 0^+} \frac{1}{x- i y}= \frac{1}{x}+i \pi \delta(x)\,
\ee
instead of (\ref{plem}), to get the right value $S^-(x)=-e^{-2 \pi s} S^+(x)$. 

In conclusion, the solutions of the problem in the complex plane (\ref{hol_A}-\ref{bc}) are in one to one correspondence with the eigenvectors of the correlator kernel (\ref{corrC}).

\subsection{Multiplicity and normalization of eigenvectors} \label{sec12}
Because of conditions (\ref{disc}) and (\ref{bc}), the function $S(z)$ must have the following asymptotic behaviour when $ z \rightarrow \partial A$,
\bea
S(z)&\sim & V_{a_i}\, (a_i-z)^{-1/2+i s}\,,\label{expa}\\
S(z)&\sim & V_{b_i}\, (z-b_i)^{-1/2-i s}\,,\label{expa2}
\eea
where $V_{a_i}$, $V_{b_i} \in \mathbb{C}$ are constants.\footnote{The difference between the left and right sides in expressions (\ref{expa}) and (\ref{expa2}) are analytic functions on $\mathbb{C} -A$ with finite limit when $z \rightarrow \partial A$.} Below, we will show these constants uniquely determine the solution. 

In order to see this, for each $s \in \mathbb{R}$ we define the Green function $G(z,w)$ for the problem (\ref{hol_A}-\ref{bc}), i.e.
\bea
& & G(z,w) \,\,\textrm{analytic on}\, w \in \mathbb{C}  - \{ w\in \bar{A}\,\textrm{or} \,\,w = z \} \, , \\
& & G(z,w) \sim (z - w)^{-1} \hspace{0.5 cm} \text{when } w \sim z \, , \\
&& \lim_{x_2\rightarrow 0^+} G(z,x_1+ i x_2)=-e^{-2 \pi s} \lim_{x_2\rightarrow 0^-} G(z,x_1+ i x_2)\,, \hspace{1cm} x_1\in A \, , \label{jumpG}
\eea
and in addition $G(z,w)$  satisfies the two boundary conditions (\ref{bc1}) and (\ref{bc}) as a function of $w$ for each $z \in \mathbb{C}$ fixed.\footnote{We explicitly change the sign of $s$ in (\ref{jumpG}) respect to (\ref{disc}).} For $w \rightarrow \partial A$ then we have an expansions analogous to (\ref{expa}),
\bea
G(z,w)& \sim &  U_{a_i}(z)\, (a_i-w)^{-1/2-i s}\, ,\label{expaG}\\
G(z,w)& \sim & U_{b_i}(z)\, (w-b_i)^{-1/2+i s}\, .
\eea
Then, the combination $G(z,w) S(w)$ does not have any jump singularity at $A$ as a function of $w$. On the other hand, it has already simple poles at $\partial A$ and at $z$, but it does not have a pole at infinity. Since the sum of all its residues must vanish, we have
\be
S(z)=\sum_{i=1}^n \left(V_{a_i} U_{a_i}(z)-  V_{b_i} U_{b_i}(z)\right) \, . \label{C2subspace} 
\ee

This shows there are at most $2 n$ linearly independent solutions to the problem (\ref{hol_A}-\ref{bc}) for fixed $s$, and they can be viewed simply as elements of $\mathbb{C}^{2n}$. It also shows that any solution which is bounded on $\partial A$ (i.e. $V_{a_i}=V_{b_i}=0$) must vanish. 

Now, we will show that the degenerancy of the space of solutions for each $s$ fixed is indeed at most $n$. Let us take two solutions $S_1(z)$ and $S_2(z)$ corresponding to the same value $s$. The function $\tilde{S}_1(z)=(S_1(z^*))^*$ is a solution with parameter $-s$ instead of $s$. The function $\tilde{S}_1(z)S_2(z)$ does not have a cut, only poles at $\partial A$. The sum of residues must vanish and we get
\be
\sum_{i=1}^n \left((V_{a_i}^1)^* V_{a_i}^2-(V_{b_i}^1)^* V_{b_i}^2\right)=0\,,\label{espacio}
\ee
where $V_{a_i}^1$, $V_{b_i}^1$ are the coefficients corresponding to $S_1 $ and $V_{a_i}^2$, $V_{b_i}^2$ the ones corresponding to $S_2$. This means that any two solutions of (\ref{hol_A}-\ref{bc}) for the same $s$ must be orthogonal according to the canonical (non positive) inner product of $\mathbb{C}^{n,n}$, which includes the case when the two solutions are the same. The argument to justify why the space of solutions is at least $n$ is as follows. Suppose that the $s$-valued subspace of solutions is spanned by $\{ S_1, \dots S_{2n} \}$, where each $S_k$ is of the form (\ref{C2subspace}). Then after a diagonalization procedure\footnote{Equivalent to the Gauss-Jordan algorithm used to diagonalize a finite dimensional matrix.} we can get a new set of solutions $\{ \tilde{S}_1, \dots \tilde{S}_{2n} \}$ which spans the same subspace but with the property that $V_{a_i}^k = 0$ for all $i=1, \dots ,n$ for all $k=n+1, \dots ,2n$. Automatically, because of (\ref{espacio}), we must have $V_{b_i}^k = 0$ for all $i=1, \dots ,n$ and for all $k=n+1, \dots ,2n$ and hence $\tilde{S}_{n+1}(z) = \dots = \tilde{S}_{2n}(z) = 0$. In conclusion, the $s$-valued subspace of solutions has dimension at most $n$. 

We will show in the next subsection that the dimension is exactly $n$. 

Now we make a final comment about the normalization of the eigenvectors $S^+(x,s)$, where we are writing explicitly the dependence of the eigenvectors through the eigenvalues $s$. Since any two eigenvectors  $S^+_1(x,s)$ and  $S^+_2(x,s')$ must be orthogonal for $s \neq s'$, we have\footnote{The function $(S^+_1(x,s))^*$ is the complex conjugate of the boundary value of $S^+_1(x,s)$ which is not the same that the boundary value of $(S_1(z,s))^*$. These two operations do not commute.}
\be
\int_A dx\,  (S^+_1(x,s))^*\,S^+_2(x,s') \propto  \delta(s-s' ) \, . \label{innerp}
\ee

In order to orthonormalize the eigenvectors, we need to figure out the proportionality constant on the above equation. For this we note the delta function can only come from the integration around the end points of the intervals on the scalar product. Using the asymptotic expansion near these points we arrive at\footnote{More precisely, we should write each eigenvector as $S(z) = \sum_{i=1}^n  V_{a_i}\, (a_i-z)^{-1/2+i s} + V_{b_i}\, (z-b_i)^{-1/2-i s} + R(z)$ where the function $R(z)$ has finite limit when $z \rightarrow \partial A$. Then after replacing in the left hand side of (\ref{innerp}) we get that the only possible delta Dirac contributions are of the form (\ref{cierto}).}
\be
\int_A dx\,  (S^+_1(x,s))^*\,S^+_2(x,s')= \pi \, e^{2 \pi s} \, \delta(s-s' )\,\sum_{i=1}^n \left( (V_{a_i}^1)^* V_{a_i}^2+(V_{b_i}^1)^* V_{b_i}^2 \right) \,.\label{cierto}
\ee 
Note the two terms inside the parenthesis in the right hand side are equal according to (\ref{espacio}).

\subsection{Construction of the eigenvectors}

In this subsection we will explicitly construct the eigenvectors of the correlator $C_A(x-y)$ using the relation developed in subsection \ref{cp}. 
Concretely, we will find the general structure of any solution $S(z)$ of the problem (\ref{hol_A}-\ref{bc}) and through them we will obtain the corresponding eigenvectors. In particular, we will show that all eigenspaces for a given eigenvalue have dimension $n$. In this subsection  $s \in \mathbb{R}$ is fixed. 

We start defining the complex valued function
\be
\tilde{\omega}(z)=\sum_{i=1}^n \log\left(\frac{z-a_i}{z-b_i}\right)\,,
\ee
where $\log$ is the principal determination of the complex logarithm which has a branch cut for $z \in \mathbb{R}_{\leq 0}$. The  function $\tilde{\omega}$ is analytic everywhere on the plane except at $\bar{A}$ where it has a jump discontinuity of the form 
\be
\tilde{\omega}^+(x)-\tilde{\omega}^-(x)=-2 \pi i\,, \hspace{1cm} x\in A\,.
\ee
Therefore, the function
\be
\mathrm{e}^{\left(i s + \frac{1}{2}\right) \tilde{\omega}(z)} \, ,
\ee
satisfies the conditions (\ref{hol_A}), (\ref{disc}) and (\ref{bc}), but it doesn't satisfy (\ref{bc1}). On the other hand, given any solution $S(z)$ of (\ref{hol_A}-\ref{bc}), the function 
\be
f(z)= S(z) \, \mathrm{e}^{-\left(i s + \frac{1}{2}\right) \tilde{\omega}(z)}
\ee
is analytic on $\mathbb{C}-\bar{A}$ and it's also continuous on $A$, and hence\footnote{By Schwartz reflection principle.} it's analytic on $\mathbb{C}-\partial A$. Then $f(z)$ must be some rational function with poles located at the end points of the intervals and also possibly at $\infty$.\footnote{A further analysis prevents the possibility of having essential singularities at such points.} 
Since $S(z)$ satisfies (\ref{bc}) and because of
\be
\lim_{z \rightarrow x + i0^+}   \left| \mathrm{e}^{\left(i s + \frac{1}{2} \right) \tilde{\omega}(z)} \right| = \mathrm{e}^{\pi s}\prod_{i=1}^n\sqrt{\left|\frac{ (x-a_i)}{(x-b_i)}\right|}\, ,\hspace{.5 cm} x\in A\,,
\ee
it follows that $f(z)$ must be of the form
\be
f(z)= \frac{g(z)}{\prod_{i=1}^n (z-a_i)}
\ee
with $g(z)$ an entire analytic function. In order to satisfy the last requirement (\ref{bc1}) for $S(z)$, we have that $g(z)$ can be a polynomial function of degree at most $n-1$. Taking all this into account, any solution $S(z)$ for the problem (\ref{hol_A}-\ref{bc}), must be of the form
\be
S(z)=\frac{\sum_{k=0}^{n-1} a_k \, z^k}{\prod_{i=1}^n (z-a_i)} \mathrm{e}^{\left(i s + \frac{1}{2} \right) \tilde{\omega}(z)} \, , \label{solut}
\ee
where $a_k \in \mathbb{C}$ parametrize $n$ linearly independent functions. Conversely, it's easy to see that any complex valued function of the form (\ref{solut}) is a solution for the problem (\ref{hol_A}-\ref{bc}).

Taking the limit of $z \rightarrow A$ from the upper side of the cut on expression (\ref{solut}), we obtain the eigenvectors
\be
S^+(x)=-i(-1)^{n-l} \mathrm{e}^{\pi s} \mathrm{e}^{i s \omega(x)} \frac{\sum_{k=0}^{n-1} a_k \, x^k}{\sqrt{- \prod_{i=1}^n (x-a_i) (x-b_i)}} \, , \hspace{.6 cm} \text{for } x\in (a_l,b_l)\,, \label{solu}
\ee
where\footnote{In \cite{modu} we have used the notation $z(x)$ for the function $\omega(x)$.}
\be
\omega(x) = \lim_{z \rightarrow x+i0^+}  \mathrm{Re} \, \tilde{\omega}(z)  = \log \left(- \frac{\prod_{i=1}^n (x-a_i)}{\prod_{i=1}^n (x-b_i)} \right) \, .
\ee
Therefore, there are exactly $n$ degenerate eigenfunctions for the same $s$. This space of eigenfunctions coincides with the one obtained in \cite{modu}.

\subsubsection{Scalar product}

Due to the degeneracy, we have some arbitrariness for the election of the eigenvectors. Such freedom is encoded in the polynomial $P(x)=\sum_{k=0}^{n-1} a_k \, x^k$ of equation (\ref{solut}). In subsection \ref{cob} we will fix such freedom in order to get an orthonormal basis of eigenvectors. In order to do that, it is useful to have an expression for the scalar product between two eigenvectors in terms of its corresponding polynomials. In the rest of this subsection, we will obtain such expression. In eq. (\ref{cierto}), using that the scalar product of two eigenfunctions is proportional to a delta function $\delta(s-s')$, we obtained these scalar products in terms of the coefficients of the expansion of the eigenvectors near the end-points of the intervals. We will re-obtain this result here by explicit integration of the product of eigenfunctions.

First we take two solutions $S^+_1(x,s)$ and $S^+_2(x,s')$ of the form (\ref{solu}) corresponding to two polynomials $P_1(x)$ and $P_2(x)$. Then, we compute the scalar product
\be
\int_A dx\, S_1^{+*}(x,s)\, S_2^+(x,s') =   -\mathrm{e}^{\pi (s+s')}\int_{-\infty}^{+\infty} d\omega\, \mathrm{e}^{-i(s-s')\omega}\sum_{l=1}^n \frac{1}{\omega'(x_l)}\frac{P_1^*(x_l) P_2(x_l)}{\prod_{i=1}^{n}(x_l-a_i)(x_l-b_i)} \, , \label{previ}
\ee
where we have changed the integration variable to $\omega$ and the sum in \eqref{previ} runs over the disitinct solutions of the equation $\omega(x)=\omega$, which are the $n$ simple roots of the polynomial equation
\be
-\mathrm{e}^\omega\prod_{i=1}^{n}(x-b_i)=\prod_{i=1}^{n}(x-a_i)\,. \label{polr}
\ee
In each interval $A_l=(a_l,b_l)$, $\omega(x)$ is a monotone increasing function that goes from $-\infty$ at $a_l$ to $+\infty$ at $b_l$.  This fact implies that there exists $n$ distinct simple roots $x_l$, each of one belonging to any distinct interval $A_l$. In \eqref{previ} $x_l$ is understood as function of $\omega$, i.e.  $x_l(\omega)$.
\\
\\
In order to proceed we will show that the following function of $\omega$
\be
K(\omega) = \sum_{l=1}^n\frac{1}{\omega'(x_l)}\frac{Q_{2n-2}(x_l)}{\prod_{i=1}^{n}(x_l-a_i)(x_l-b_i)} \,,\label{constant}
\ee
is a constant, i.e. $K(\omega)$ is independent of $\omega$ for any polynomial $Q_{2n-2}(x)$ of degree $2n-2$.
Replacing the following expression for $\omega'(x)$
\be
\omega'(x)=\frac{\prod_{i=1}^{n}(x-b_i)\sum_{k=1}^n\prod_{j\neq k}(x-a_j)-\prod_{i=1}^{n}(x-a_i)\sum_{k=1}^n\prod_{j\neq k}(x-b_j)}{\prod_{i=1}^{n}(x-a_i)(x-b_i)} \, , 
\ee
in \eqref{constant}, we arrive at 
\be
K(\omega) = \sum_{l=1}^n \frac{Q_{2n-2}(x_l)}{\prod_{i=1}^{n}(x_l-b_i)\sum_{k=1}^n\prod_{j\neq k}(x_l-a_j)-\prod_{i=1}^{n}(x_l-a_i)\sum_{k=1}^n\prod_{j\neq k}(x_l-b_j)}. \label{into}
\ee
Since $\omega=-\infty$ implies $x_l=a_l$ and $\omega=+\infty$ implies $x_l=b_l$,  then we have the following particular limits
\bea
K(-\infty) &= & \sum_{l=1}^n \frac{Q_{2n-2}(a_l)}{\prod_{i=1}^{n}(a_l-b_i)\prod_{j\neq l}(a_l-a_j)}\,,\label{dio}\\
K(\infty) & = & -\sum_{l=1}^n \frac{Q_{2n-2}(b_l)}{\prod_{i=1}^{n}(b_l-a_i)\prod_{j\neq l}(b_l-b_j)}\,.\label{dio1}
\eea
Now, we will show $K(\omega)=K(-\infty)$, and hence constant. For this, from equation \eqref{polr} we have the following polynomial identity
\be
\mathrm{e}^\omega\prod_{i=1}^{n}(x-b_i)+\prod_{i=1}^{n}(x-a_i)=(e^{\omega}+1)\prod_{l=1}^n (x-x_l)\, . \label{yyy}
\ee
Evaluating \eqref{yyy} on $x=a_k$ (for some $k=1,\cdots,n$) we get
\be
\prod_{i=1}^n (a_k-b_i)=(1+\mathrm{e}^{-\omega}) \prod_{l=1}^n (a_k-x_l)  \,,\label{refe}
\ee
and taking the derivative of \eqref{yyy} respect to $x$ and evaluating at $x=x_l$ (for some $l=1,\cdots,n$) we have
\be
\prod_{i=1}^{n}(x_l-b_i)\sum_{k=1}^n\prod_{j\neq k}^{n}(x_l-a_j)-\prod_{i=1}^{n}(x_l-a_i)\sum_{k=1}^n\prod_{j\neq k}^{n}(x_l-b_j)=-(1+\mathrm{e}^{-\omega}) \prod_{i=1}^n (x_l-a_i)\prod_{j\neq l}^n (x_l-x_j)\,. \label{refe2}
\ee
Then, replacing \eqref{refe} on the denominator of \eqref{dio} we get
\be
K(-\infty) = (1+\mathrm{e}^{-\omega})^{-1}  \sum_{l=1}^n \frac{Q_{2n-2}(a_l)}{\prod_{i=1}^{n}(a_l-x_i)\prod_{j\neq l}(a_l-a_j)}\,,
\ee
and replacing \eqref{refe2} on the denominator of \eqref{into} it follows 
\be
K(\omega)= - (1+\mathrm{e}^{-\omega})^{-1} \sum_{l=1}^n \frac{Q_{2n-2}(x_l)}{\prod_{i=1}^n (x_l-a_i)\prod_{j\neq l}^n (x_l-x_j)}.
\ee
Hence, the expected relation $K(\omega)= K(-\infty)$ follows from
\be
(1+\mathrm{e}^{-\omega}) \left[ K(-\infty) - K(\omega) \right]= \sum_{l=1}^n \frac{Q_{2n-2}(x_l)}{\prod_{i=1}^n (x_l-a_i)\prod_{j\neq l}(x_l-x_j)}+\sum_{l=1}^n \frac{Q_{2n-2}(a_l)}{\prod_{i=1}^n (a_l-x_i)\prod_{j\neq l}(a_l-a_j)} = 0\, , \label{ppp}
\ee
where the last equality to zero is a general fact valid for any polynomial $Q_{2n-2}$ of degree $2n-2$: evaluating the polynomial in $2n$ arbitrary points $y_1,\cdots,y_{2n}$ there is a linear equation that relates the value on the first $2n-1$ points to the value on $y_{2n}$. This equation is 
\be
\sum_{l=1}^{2n} \frac{Q_{2n-2}(y_l)}{\prod_{i\neq l} (y_l-y_i)}=0\,.  
\ee
Eq. (\ref{ppp}) follows specializing on $y_i=a_i$ (for $i=1,\cdots, n$), and  $y_i=x_{i-n}$ (for $i=n+1,\cdots,2n$). 
\\
\\
Since $K(\omega)$ is constant, we have that $K(-\infty)=K(\infty)$, i.e. expressions \eqref{dio} and \eqref{dio1} are the same. This in fact coincides with the relation \eqref{espacio} for the coefficients of the expansions \eqref{expa} and \eqref{expa2}  for the solutions \eqref{solut} at the end points of the intervals.  Reading off these coefficients from the explicit form of the solutions, the relation (\ref{espacio}) writes
\bea 
\sum_{i=1}^n (V^1_{a_i})^* V^2_{a_i} &=& - 
\sum_{l=1}^n \frac{P_1(a_l)^*P_2(a_l)}{\prod_{i=1}^{n}(a_l-b_i)\prod_{j\neq l}(a_l-a_j)} \nonumber \\
&=&\sum_{l=1}^n \frac{P_{1}(b_l)^* P_2(b_l)}{\prod_{i=1}^{n}(b_l-a_i)\prod_{j\neq l}(b_l-b_j)} = \sum_{i=1}^n (V^1_{b_i})^*V^2_{b_i}\,. \label{vvz}
\eea 
\\
\\
Let us come back to the scalar product \eqref{previ}. Since the integrand on the Fourier transform in $\omega$ is constant, we get
\be
\int_A dx\, S_1^{+*}(x,s)\, S_2^+(x,s')= -2\pi \, e^{2 \pi s} \, \delta(s-s') \sum_{l=1}^n \frac{P_1(a_l)^* P_2(a_l)}{\prod_{i=1}^{n}(a_l-b_i)\prod_{j\neq l}(a_l-a_j)} \, ,\label{escalaa}
\ee
which coincides with the equation \eqref{cierto} because of \eqref{vvz}.

\subsection{A complete orthonormal basis} \label{cob}

In order to construct a basis of eigenvectors for each eigenspace of fixed $s$, we choose the following subset  $ \{ u^k_s \}_{k=1}^n$ of eigenfunctions
\be
u_s^k(x)=\frac{(-1)^{l+1}}{N_k}\frac{\prod_{i\neq k}(x-a_i)}{\sqrt{-\prod_{i=1}^n(x-a_i)(x-b_i)}}\mathrm{e}^{i s \omega(x)}\,,\hspace{.7cm} x\in (a_l,b_l)\,,  \label{solunorm}
\ee
with the normalization factor\footnote{Note that the expression apparently differs from eq. (36) of \cite{modu}. But
\be
\sum_{j=1}^n \frac{\prod_{l\neq k}(b_j-a_l)}{(b_j-a_k)\prod_{l\neq j}(b_j-b_l)}=\frac{\prod_{i\neq k}(a_i-a_k)}{\prod_{i=1}^n (b_i-a_k)}\,,
\ee
and then both equations are in agreement.}
\be
N_k=\sqrt{2\pi}\left(\frac{\prod_{i\neq k}(a_i-a_k)}{\prod_{i=1}^n (b_i -a_k )} \right)^{1/2}\,. \label{Nk}
\ee
In the rest of this subsection, we will show that the set $ \{ u^k_s\}_{k=1}^n$ is orthonormal and complete.

The orthonormality follows immediately from equation (\ref{escalaa}), and hence we have
\be
\int_A dx \, u_s^{k*}(x)u_{s'}^{k'}(x)=\delta_{k,k'}\,\delta(s-s')\, .
\ee

The completeness is quite less obvious. The general argument of section \ref{sec12} shows that $n$ is the maximal degeneracy and then any $n$ linearly independent vectors should form a complete basis. This fact can be shown explicitly as follows. 

Using the eigenfunctions \eqref{solunorm}, we have 
\be
\sum_{k=1}^n \int_{-\infty}^{\infty} ds\, u_s^{k}(x)u_s^{k*}(y)=2 \pi \, \sqrt{\prod_{i=1}^n\frac{(x-a_i)(y-a_i)}{(x-b_i)(y-b_i)}}\left( \sum_{k=1}^n\frac{1}{N_k^2(x-a_k)(y-a_k)} \right) \sum_{l=1}^n \frac{1}{\omega'(x_l)} \delta(x-x_l) \, ,\label{dels}
\ee
where $x_l\equiv x_l(\omega(y)) \in A_l$ are the $n$ roots of the polynomial equation \eqref{polr} for  $\omega=\omega(y)$. In particular, when $y \in A_l$ then $x_l \equiv y$.
Using the following algebraic relation\footnote{This relation is true for any complex numbers $a_1,\cdots,a_n,b_1,\cdots,b_n,x,y$. It can be proven using the definition \eqref{Nk} for the normalization constants $N_k$ and decomposing the rational function at both sides into the poles for the variable $x$.}
\bea
\sum_{k=1}^n \frac{1}{N_k^2(x-a_k)(y-a_k)}&=&\frac{\prod_{k=1}^n(x-b_k)(y-a_k)-\prod_{k=1}^n(y-b_k)(x-a_k)}{2\pi(x-y)\prod_{k=1}^n (x-a_k)(y-a_k)} \nonumber \\ 
&=&\frac{P(x,y)}{2\pi\prod_{k=1}^n(x-a_k)(y-a_k)}\,,\label{trick}
\eea
where the function
\be
P(x,y)=\frac{\prod_{k=1}^n(x-b_k)(y-a_k)-\prod_{k=1}^n(y-b_k)(x-a_k)}{x-y}\,,\label{sx}
\ee 
is a polynomial in $x$ of degree $n-1$ for each fixed $y$, and its roots are the points $x=x_l$ except when $x_l=y$. Because of that, the only delta function which survives in \eqref{dels} is for $x=y$ and hence
\be
\sum_{k=1}^n \int_{-\infty}^{\infty} ds\, u_s^{k}(x)u_s^{k*}(y)= - \frac{P(x,x)}{\prod_{k=1}^n(x-a_k)(x-b_k)} \frac{1}{\omega'(x)}\delta(x-y) \, . \label{peda}
\ee
In order get a better expression for $P(x,x)$,  from \eqref{sx} we define a new function
\be
Q(x,y)= P(x,y) (x-y) = \prod_{k=1}^n(x-b_k)(y-a_k)-\prod_{k=1}^n(y-b_k)(x-a_k) \, ,
\ee
which allows us to compute
\be
P(x,x)=\partial_x Q(x,y)|_{y=x}= - \omega'(x) \prod_{k=1}^n(x-a_k)(x-b_k)\, .\label{zo}
\ee
Finally, replacing \eqref{zo} into \eqref{peda} we obtain the completeness relation 
\be
\sum_{k=1}^n \int_{-\infty}^{\infty} ds\, u_s^{k}(x)u_s^{k*}(y) =  \delta(x-y) \, .
\ee

\subsection{Modular Hamiltonian}
In this subsection we re-derive the results of \cite{modu} about the modular Hamiltonian using the information about the spectral decomposition of the correlator kernel $C_A(x-y)$ obtained in the previous subsections. The modular flow corresponding to this modular Hamiltonian and the entanglement entropy for several intervals have been computed in \cite{modu}. Recently, the modular Hamiltonian has also been computed using Euclidean path integral methods in \cite{wong}. In \cite{rehren} it was shown that the modular flow satisfies the KMS condition. In \cite{longo} the mutual information between several intevals have been computed using the Araki formula without using a cutoff to compute the entanglement entropy. The results coincide with the ones in \cite{modu}. 

From \eqref{coco} and \eqref{lami} the correlator kernel writes
\be
C_A(x-y) = \sum_{k=1}^n \int_{-\infty}^{+\infty} ds\,  u_s^k(x)  \frac{1+ \tanh(\pi s)}{2} u_s^{k\, *}(y) \, , 
\ee
and using this formula and \eqref{hache} we obtain the following expression for the modular Hamiltonian kernel
\be
H(x,y) = \sum_{k=1}^n \int_{-\infty}^{+\infty} ds \, u_s^k(x) \, 2 \pi s \, u_s^{k\, *}(y) \, . \label{H_ker} 
\ee
Using equation \eqref{solunorm} we get 
\be
H(x,y)  = -i 2 \pi \, k(x,y) \, \delta'\left(\omega\left(x\right)-\omega\left(y\right)\right) \, , \\ \label{H_ker2} 
\ee
where the function
\be
k(x,y)  = 2\pi\sqrt{\prod_{i=1}^{n}\frac{\left(x-a_{i}\right)\left(y-a_{i}\right)}{\left(x-b_{i}\right)\left(y-b_{i}\right)}}\left(\sum_{k=1}^{n}\frac{1}{N_{k}^{2}\left(x-a_{k}\right)\left(y-a_{k}\right)}\right) \, . \label{defK}
\ee
 The aim for the rest of this subsection, is to obtain a simplified expression for the modular Hamiltonian kernel  \eqref{H_ker2}. First we have the following identity for the Dirac delta term
\be
\delta'\left(\omega\left(x\right)-\omega\left(y\right)\right)  =  \sum_{l=1}^{n}\delta'\left(x-x_{l}\right)\frac{1}{\omega'\left(x\right)^{2}}-\delta\left(x-x_{l}\right)\frac{\omega''\left(x\right)}{\omega'\left(x\right)^{3}} \, ,  \label{deltacomp}
\ee
where $x_l \equiv x_l(y) \in A_l$ are the roots of $\omega(x)=\omega(y)$ introduced in equation \eqref{dels}. From this last equation, the modular Hamiltonian splits into the sum of a local and a non-local operators
\be
H(x,y) = H_{loc}(x,y) + H_{noloc}(x,y) \, ,
\ee
where $H_{loc}(x,y)$ comes from the term in \eqref{deltacomp} when $x_l(y)=y$ and $H_{noloc}(x,y)$ comes from the $n-1$ terms in \eqref{deltacomp} when $x_l(y) \neq y$. We discussed these two contributions separately.

\subsubsection{Local part}

We recognize the local part for the modular Hamiltonian kernel as
\be
H_{loc}(x,y) = -i 2 \pi \, k(x,y) \,  \left[ \frac{1}{\omega'(x)^2} \delta'(x-y) - \frac{\omega''(x)}{\omega'(x)^3} \delta(x-y) \right] \, . \label{Hloc} 
\ee
In order to simplify the above expression, it is compulsory to understand it as a distribution acting over some smooth compactly supported test function $\varphi(x,y)$. Integrating by parts, the derivative of the Dirac delta is converted to
\bea
\varphi(x,y)\, k(x,y) \, \frac{1}{\omega'(x)^2} \delta'(x-y) &=& - \Bigg[ \partial_x \varphi(x,y) \, k(x,x) \, \frac{1}{\omega'(x)^2} + \varphi(x,x)\, \left. \partial_x k(x,y) \right|_{y=x}  \frac{1}{\omega'(x)^2} \nonumber \\
& & + \varphi (x,y)\, k(x,x) \, \partial_x \left( \frac{1}{\omega'(x)^2} \right) \Bigg] \delta(x-y) \, , \label{deltap}
\eea
which can be simplified after we recognize the following identities
\bea
k(x,x) & = & \omega'(x) \, , \label{kxx} \\
\left. \partial_x k(x,y) \right|_{y=x} & = & \frac{1}{2}\omega''(x) \, .  \label{dkxx}
\eea
which they follow from \eqref{defK} and the algebraic relation\footnote{As \eqref{trick}, equation \ref{trick2} is a pure algebraic relation valid for any complex number $a_k$, $b_k$ and $x$. It can be shown matching the coefficients of the poles in $x$ on both sides.}
\be
\sum_{k=1}^{n}\frac{1}{N_k^2}\frac{1}{(x-a_k)}  =  \frac{1}{2\pi}\left(1+\mathrm{e}^{-\omega(x)}\right) \, . \label{trick2}
\ee
The final steps consist on replacing \eqref{kxx} and \eqref{dkxx} into \eqref{deltap}, and integrating by parts the term containing $ \partial_x \varphi(x,y)$, in order to factorize the test function. 
 We get
\be
H_{loc}(x,y) = -i 2 \pi \left[ \frac{1}{\omega'(x)} \delta'(x-y) + \frac{1}{2} \partial_x \left( \frac{1}{\omega'(x)} \right) \delta(x-y)  \right] \, . \label{Hlocf}
\ee
The local part of the modular Hamiltonian comes from \eqref{citric} and \eqref{Hlocf}
\be
\mathcal{H}_{loc} = 2\pi\int_{A}dx\,\frac{1}{\omega'(x)}T(x) \, ,
\ee
where $ T(x) = \frac{1}{2}\left[i \partial_x \psi^\dagger(x) \psi(x)- \psi^\dagger (x) i \partial_x \psi(x) \right]$ is the energy density operator.

\subsubsection{Non local part}

The non local part of the modular Hamiltonian kernel is
\be
H_{nonloc}(x,y) = -i 2 \pi \, k(x,y) \,  \left[ \sum_{l=1\, x_l \neq y}^n \frac{1}{\omega'(x)^2} \delta'(x-x_l) - \frac{\omega''(x)}{\omega'(x)^3} \delta(x-x_l) \right] \, . \label{Hnonloc} 
\ee
The first term can be simplified by a similar computation as we did around eq. \eqref{deltap}. Here the situation is simpler because $k(x_l,y)\equiv 0$ for all $x_l \neq y$ as we showed in \eqref{trick}. Hence the unique term which survives it is the one proportional to de derivative of $k(x,y)$,
\be
\left. \partial_x k(x,y) \right|_{x=x_l} = \frac{\omega'(x_l)}{x_l -y} \, .
\ee
Replacing it on \eqref{Hnonloc} we arrive to
\be
H_{noloc}(x,y) = i 2 \pi \sum_{l=1,\,x_l \neq y}^n \frac{1}{(x-y)} \frac{1}{\omega'(x)} \delta \left(x-x_l\left(\omega\left(y\right)\right)\right) \, , \label{kernelnl}
\ee
and 
\bea
\mathcal{H}_{noloc}  & = & i\,2\pi\sum_{l=1,\,x_{l}\neq x}^{n}\int_{A}dx\,\psi^{\dagger}\left(x_{l}\right)\frac{1}{\left(x_{l}-x\right)}\frac{1}{\omega'\left(x_{l}\right)}\psi\left(x\right)\\
 & = & i\,2\pi\int_{A\times A,\,x\neq y}dx\,dy\,\psi^{\dagger}\left(x\right)\frac{\delta\left(\omega\left(x\right)-\omega\left(y\right)\right)}{x-y}\psi\left(y\right) \, .
\eea

\subsubsection{Two intervals}

For the case of two intervals $A= (a_1,b_1) \cup (a_2,b_2)$, the modular Hamiltonian operator $ \mathcal{H} = \mathcal{H}_{loc}+\mathcal{H}_{noloc}$ reduces to

\bea
{\cal H} &=& 2\pi\int_{A}dx\, \omega'(x)^{-1}\,T\left(x\right)  + i2\pi\int_{A}dx\,\psi^{\dagger}(x)\omega'(x)^{-2}\frac{\left(b_{1}-a_{1}\right)\left(a_{2}-b_{1}\right)\left(b_{2}-a_{1}\right)\left(b_{2}-a_{2}\right)}{\left(x-a_{1}\right)\left(x-a_{2}\right)\left(x-b_{1}\right)\left(x-b_{2}\right)} \nonumber \\
&& \frac{1}{x\left(a_{1}+a_{2}-b_{1}-b_{2}\right)+\left(b_{1}b_{2}-a_{1}a_{2}\right)}\psi\left(\bar{x}\right) \, ,
\eea 
where
\bea
\omega'(x) &=& \frac{1}{x-a_{1}}+\frac{1}{x-a_{2}}-\frac{1}{x-b_{1}}-\frac{1}{x-b_{2}} \, , \\
\bar{x} &=&\frac{a_{1}a_{2}\left(x-b_{1}-b_{2}\right)-b_{1}b_{2}\left(x-a_{1}-a_{2}\right)}{x\left(a_{1}+a_{2}-b_{1}-b_{2}\right)+\left(b_{1}b_{2}-a_{1}a_{2}\right)} \, . \label{conj-points}
\eea

For two symmetric intervals $A=(-R,-r)\cup (r,R)$, $0<r<R$, a more explicit form of ${\cal H}$ is 
\be
{\cal H} = 2\pi\int_{A}dx\,\frac{\left(x^{2}-r^{2}\right)\left(R^{2}-x^{2}\right)}{2\left(R-r\right)\left(x^{2}+rR\right)}\,T\left(x\right) + i\pi\int_{A}dx\,\psi^{\dagger}\left(x\right)\frac{rR\left(x^{2}-r^{2}\right)\left(R^{2}-x^{2}\right)}{\left(R-r\right)x\left(x^{2}+rR\right)^{2}}\psi\left(\bar{x}\right) \, , 
\ee 
where now $\bar{x}= - \frac{r R}{x}$.

\section{The free chiral current}
\label{escalar}
We are going to study the model  determined by a field operator $j(x)$ in the line, which is identified with the chiral derivative of a massless free scalar in $d=2$, that is, $j(x^+)=\partial_+ \phi(x^+)$, with $x^+=x^0+x^1$. Then the line we are considering can be thought as a null line in the $d=2$ model, and all variables $x$, $y$, etc. will be null variables. We will study the structure of the vacuum density matrix reduced to a region. Because of the complexity of the problem we are going to restrict attention to the case of one or two intervals.
 
The commutator is
\be
[j(x),j(y)]=i\, \delta'(x-y)\,,\label{commut}
\ee
and the model has Hamiltonian
\be
H=\frac{1}{2}\int dx\, j^2(x)\,,
\ee
where $T(x)=(1/2)j^2(x)$ is the energy density operator. 

The vacuum two point correlator distribution is 
\be
F(x,y)=\langle j(x) j(y)\rangle=-\frac{1}{2\pi}\frac{1}{(x-y-i\, 0^+)^2}\,.\label{cotto}
\ee
The model is Gaussian and all other multipoint correlators follow from this one according to Wick's theorem.

In order to proceed we will need general formulae for Gaussian states in an algebra of canonical commutation relations. We briefly review the derivation of these formulas in the appendix \ref{modularscalar}.  
 
An algebra of canonical commutation relations can be written in the form
\be
\left[f_i,f_j\right]= i C_{ij}\,,
\ee
for hermitian variables $f_i$, with numerical commutator given by the real antisymmetric matrix $C_{ij}$. We take a  Gaussian state with correlator
\be
F_{ij}=\langle f_i f_j\rangle\,.
\ee
 The modular Hamiltonian is then given by 
\bea
{\cal H} &=&\sum_{ij} f_i H_{ij} f_j\,,\\
H &=&  -\frac{i}{2} \,  \frac{V}{|V|}\log\left(\frac{|V|+1/2}{|V|-1/2}\right) \,C^{-1} \,,\label{hamidos}
\eea
where
\be
V=-i C^{-1} F-\frac{1}{2}\,.\label{dit2}
\ee
The entropy is \cite{sorkin}
\be
S=\textrm{tr}\,  \Theta(V)\,\left((V+1/2)\log(V+1/2)+(1/2-V)\log(V-1/2)\right)\,. \label{39}
\ee

The operator $V$ is not symmetric but it has real eigenvalues $\nu \in \pm(1/2,\infty)$. In the present case this spectrum will be continuous.  For later convenience we parametrize
\be
\nu=\frac{1}{2}\coth(\pi s)\,,\hspace{1cm}s\in R\,.
\ee
 We name the left and right eigenvectors of $V$ as
\bea
V |u_s^k \rangle &=& \frac{1}{2}\coth(\pi s) |u_s^k \rangle\,, \label{dit3}\\
V^\dagger |v_s^k\rangle &=& \frac{1}{2}\coth(\pi s) |v_s^k \rangle\,,\label{dit1} 
\eea
where $k$ is a possible degeneracy index. 
We normalize the eigenvectors as  
\be
\langle v_s^k|u_{s'}^{k'}\rangle=\delta_{k,k'}\,\delta(s-s')\,.\label{manja}
\ee

It is not difficult to see from (\ref{dit2}), (\ref{dit3}), (\ref{dit1}), that $C |u_s^k\rangle$ is an eigenvector of $V^\dagger$ with the same eigenvalue as the vectors $|v_s^k\rangle$, and then is a linear combination of these later. The orthogonality (\ref{manja}) leave us the freedom to redefine the basis $|u_s^k\rangle$ by an arbitrary matrix and the basis  $|v_s^k\rangle$ by the inverse adjoint matrix. We can use this freedom to set $|v_s^k\rangle$ proportional to $C\, |u_s^k\rangle$. The phase of the proportionality constant is however fixed to be $i \,\textrm{sign}(s)$, as can be seen from taking the scalar product of (\ref{dit3}) with $\langle v_{s'}^{k'}|$ and using (\ref{manja}) and the positivity of $F$. As a result, we can further fix the vectors by taking
\be
|v_s^k\rangle= i\, \textrm{sign}(s)\, C |u_s^k\rangle\,.\label{cuen}
\ee

In terms of these vectors the kernel (\ref{hamidos}) writes simply
\be
H=\sum_k \int_{-\infty}^{\infty} ds\, |u^k_s\rangle \, \pi |s|\, \langle u^k_s| \,.\label{hkernel}
\ee

\section{The case of a single interval}
\label{unintervalo}
Let us first consider the simplest case of a single interval $A=(a,b)$. The inverse of the commutator  $\delta'(x-y)$ acting on a test function $h(x)$ has to be a linear combination of  $\int_{a}^x dy\, h(y)$, and $\int_{a}^{b} dy\, l(y)\, h(y)$, for some $l(y)$. This last term is linear in $h(y)$ and being independent of $x$, is annihilated by the $\delta'$.  
In kernel notation we have to combine\footnote{Only two of these three kernels are linearly independent.} 
\be
\theta(x-y)\,, \hspace{.5cm}-\theta(y-x)\,,\hspace{.5cm} l(y)\,.
\ee
There is only one antisymmetric inverse, given by  
\be
(\delta')^{-1}(x,y)=\frac{1}{2}\left(\theta(x-y)-\theta(y-x)\right)\,.\label{dit}
\ee
On a test function its action is
\be
((\delta')^{-1} h)(x)=\frac{1}{2}\left(\int_a^x dy\, h(y)-\int _x^b dy\, h(y)\right)\,.    
\ee
Hence $\delta'\cdot(\delta')^{-1}=1$ and $(\delta')^{-1}\cdot \delta'=1$ on test functions that vanish on the boundary of the interval.

Hence our first task is to solve the spectrum of 
\be
2 \pi C^{-1}F\equiv \frac{1}{(x-y-i\, 0^+)} -\frac{1/2}{(b-y)}-\frac{1/2}{(a-y)}\,.
\label{dospi}\ee
We proceed as in the case of the fermion. 
We think the interval $A$ as included in the real axis of the complex plane. We consider an analytic function $S(z)$, with a multiplicative boundary condition on the interval $A$, as in (\ref{disc}). This is 
\bea
&& S(z) \text{ analytic in }\mathbb{C}-\bar{A} \, , \label{hol_A1} \\
&& S^{+}(x_1)=\lim_{x_2\rightarrow 0^+} S(x_1+ i x_2)=\lambda \lim_{x_2\rightarrow 0^-} S(x_1+ i x_2)=\lambda\, S^-(x_1) \, , \hspace{1cm} x_1\in A \,. \label{disc1} 
\eea
Thus, $S(z)$ has a cut over $A$ with multiplicative boundary conditions.
We further impose the boundary conditions at the infinity and at the end points of the interval,
\bea
&& \lim_{z\rightarrow \infty} \left| \, S(z) \right| <\infty\,, \label{bc11} \\
&& \lim_{z \rightarrow \partial A} |S(z)| <\infty \, .\label{bc22}
\eea
 Consider now the complex integral
\be
\oint dz_2 \, \left(\frac{1}{z_1-z_2}-\frac{1}{2}\frac{1}{b-z_2} -\frac{1}{2}\frac{1}{a-z_2}\right)S(z_2)=0\,  \label{circint1}
\ee
on a closed curve in the complex plane encircling both $A$ and $z_1$. This integral is zero because the integrand has zero residue at infinity. We can shrink the integration contour around the point $z_1$ and around the cut to get  
 \be
 S(z_1)=\frac{i}{2 \pi } (1-\lambda^{-1}) \fint_A dy\,  \left(\frac{1}{z_1-y}-\frac{1}{2}\frac{1}{b-y} -\frac{1}{2}\frac{1}{a-y}\right) S^+(y)\,.\label{siosi}
\ee
The symbol $\fint$ for the integral means here that it is regularized at the end points of the interval by taking the complex integral on a small circle around these end-points (as implied by (\ref{circint1})) and then take the limit of zero circle size. We will soon be more specific on how this regularization can be expressed directly for the function of a real variable such as $S^+(y)$. 

Taking the limit of $z_1\rightarrow x-i 0^+$,  $x\in A$, and using (\ref{disc1}), (\ref{dospi}) and (\ref{dit2}), we get
\be
\fint_A dy\, V(x,y) \, S^{+}(y)=\frac{\lambda+1}{2(1-\lambda)} \, S^+(x)\,.\label{rumania}
\ee
Since the eigenvalues of $|V|$ are in $(1/2,\infty)$, we have $\lambda>0$, in contrast to the case of the fermion where the factor $\lambda$ is negative. We write 
\be
\lambda=e^{-2 \pi s}\,,\hspace{1cm}s\in R\,.
\ee
 The eigenvalue in (\ref{rumania}) then coincides with $1/2 \coth(\pi s)$ as in (\ref{dit3}). 

Therefore, for each solution $S(z)$ of the problem in the complex plane we get an eigenvector of the kernel $V(x,y)$ in the interval. The eigenvector modulus, in contrast to the case of the fermion, is bounded at the end points of the interval, (\ref{bc22}). This is in accordance with boundary conditions for scalars \cite{review}. Conversely, if we have a solution of (\ref{rumania}) we can use it as boundary data on the interval in (\ref{siosi}), which gives a solution $S(z)$ satisfying  (\ref{hol_A1}--\ref{bc22}). These problems are then mutually equivalent.

For a single interval we can write a solution to this problem as 
\be
S(z)=e^{-i s  \tilde{\omega}(z)}\,, \hspace{1cm} \tilde{\omega}(z)= \log\left(\frac{z-a}{z-b}\right)\,,\hspace{1cm}
\tilde{\omega}^+(x)-\tilde{\omega}^-(x)=-2 \pi i\,, \hspace{1cm} x\in A\,.\label{yy}
\ee
This obeys all boundary conditions. Any other solution divided by $S(z)$ in (\ref{yy}) must be an analytic function on the plane except perhaps the end points of the interval. Consequently this other solution would be either proportional to (\ref{yy}) or it is not bounded at infinity or at the end points of the interval. Hence (\ref{yy}) is in fact the unique solution to the problem. 

The eigenvector is given by the boundary value on the interval
\be
u(x)\propto S^+(x)\propto e^{-i s \omega(x)}\,, \hspace{1cm} \omega(x)=\log\left(\frac{x-a}{b-x}\right)\,. 
\ee

Now we explain more precisely the regularization in (\ref{siosi}), (\ref{rumania}). Frequently we will encounter integrals on the real line of the form 
\be
\int_a^b dx\, f(x)   
\ee
where 
\be
f(x)\sim c\, \frac{e^{-i s \log(x-a)}}{x-a}\,, \hspace{1cm} x\rightarrow  a\,.
\ee
Then the integral will have an oscillatory but bounded term $-c\,e^{-i s \log(x-a)}/(-i s)$ in the lower boundary  as $x\rightarrow a$,  and it does not converge. The regularization used above just subtracts this oscillatory phase, that is
\be
\fint_a^b dx\, f(x)=\lim_{\epsilon\rightarrow 0} \int_{a+\epsilon}^b dx\, f(x) + c\frac{e^{-i s \epsilon}}{(-i s)}= \int_a^b dx\, \left(f(x)- c\, \frac{e^{-i s \log(x-a)}}{x-a}\right)+ c\, \frac{e^{-i s \log(b-a)}}{(-i s)}\,. \label{fint_reg}
\ee 
If the oscillatory term appears on the upper end of the integral an analogous subtraction is understood. As mentioned above, this subtraction appears naturally when the integral comes from a limit of a complex integral around the cut as in the transformation from (\ref{circint1}) to (\ref{siosi}). The definition of the kernel $V$ has to be understood with this regularization.\footnote{Note that this regularization eliminates from the bare integral an infinitely oscillatory phase in $s$, which will produce vanishing terms in any finite calculation involving integrals over the eigenvalues.}

Now we have to look at the left eigenvectors of $V$.   These are eigenvectors of $V^\dagger=-i F C^{-1}-1/2$. For this we can just use the relation (\ref{cuen}). However, we find instructive to compute them directly from the kernel. From  (\ref{dit}) this is 
\be
2 \pi F C^{-1}(x,y)= \frac{1}{(x-y-i\, 0^+)} -\frac{1/2}{(b-x)}-\frac{1/2}{(a-x)}\,.
\ee
Now we take a new analytic function $S(z)$, and assume the same multiplicative boundary condition (\ref{disc1}). However, in order to obtain a solution of the eigenvector problem from the complex integral, we are now forced to impose $S(z)\rightarrow |z|^{-2}$ at infinity, and that $S(z)$ must have at most pole singularities at $a$ and $b$. We 
integrate
\be
\left(\frac{1}{z_1-z_2}-\frac{1}{2}\frac{1}{b-z_1} -\frac{1}{2}\frac{1}{a-z_1}\right)S(z_2)
\ee
in a close contour encircling $z_1$ and the cut $A$.  We get
\be
 S(z_1)=\frac{i}{2 \pi } (1-\lambda^{-1}) \fint_A dy\,  \left(\frac{1}{z_1-y}-\frac{1}{2}\frac{1}{b-z_1} -\frac{1}{2}\frac{1}{a-z_1}\right) S^+(y)\,.
\ee
The limit $z_1\rightarrow x-i 0^+$, $x\in A$, gives
\be
\fint_A dy\, V^\dagger(x,y) \, S^{+}(y)=\frac{\lambda+1}{2(1-\lambda)} \, S^+(x)\,.
\ee

The value of $\lambda$ is the same as above, giving the same multiplicative boundary conditions for $S(z)$ as for the eigenvectors of $V$. However,  the boundary conditions at infinity and at the end points of the interval are different. These now imply that the unique solution is
\be
S(z)=e^{-i s  \tilde{w}(z)}\left(\frac{1}{z-a}-\frac{1}{z-b}\right)\,.
\ee
The poles have to have opposite sign in order that the function decays at infinity as $|z|^{-2}$.
We recognize this function is proportional to the derivative of (\ref{yy}), as it must be, given (\ref{cuen}). 

Orthonormalizing the eigenvectors we get
\bea
u_s &=& \frac{e^{- i s  \omega(x)}}{\sqrt{2 \pi |s|}}\,,\\
v_s &=& i \,\textrm{sign}(s)\, u_s'=\sqrt{\frac{|s|}{2 \pi}}\,e^{-i s \omega(x)}\left(\frac{1}{x-a}-\frac{1}{x-b}\right)\,. \label{up}
\eea
These satisfy (\ref{manja}) and (\ref{cuen}).

\subsection{Modular Hamiltonian and entropy}

Replacing these formulas for the eigenvectors into the equation (\ref{hkernel}) and after a simple integration, we get the following expression for the modular Hamiltonian kernel 
\be
H(x,y)=\int_{-\infty}^\infty ds\, u_s(x) \, \pi |s|\, (u_s(y))^*=\pi (\omega'(x))^{-1} \,\delta(x-y)\,.  \label{mh}
\ee
Then the modular Hamiltonian operator has the known form for an interval in a CFT \cite{spacetime2}
\be
{\cal H} = 2 \pi \int_a^b dx\,  \frac{(b-x)(x-a)}{b-a}\,    T(x)\,,
\ee
where in the present case the energy density is $T=\frac{1}{2} \, j^2$.

According to (\ref{39}) the entropy is
\be
S=\int_{0}^\infty ds\, g(s) \int_A dx\, \, u_s(x) (v_s(x))^*\,, \label{s1int}
\ee
where
 \be
g(s)=\frac{1+\coth\left(\pi s\right)}{2}\log\left(\frac{\coth\left(\pi s\right)+1}{2}\right)+\frac{1-\coth\left(\pi s\right)}{2}\log\left(\frac{\coth\left(\pi s\right)-1}{2}\right) \, . \label{gs}
\ee 
The full integral over the $x$ coordinate gives a delta function $\delta(0)$ and is divergent. This is just the usual divergence of entropy in QFT due to the continuum spectrum of the modular operator. 

A convenient way to regularize the entropy is to integrate up to a distance cutoff from the boundary. 
Then we compute
\be
S = \int_{a+\epsilon}^{b-\epsilon} dx\,  \int_{0}^{\infty} ds\, g(s)\,\,u_s(x) (v_s(x))^* =\frac{1}{12}\int_{a+\epsilon}^{b-\epsilon} dx\,\omega(x)' 
= \frac{1}{6}\log\left(\frac{b-a}{\epsilon}\right)\,. \label{s1}
\ee
This gives the expected result for a conformal model with one chiral component of central charge $c=1$.
 
 According to (\ref{renyi4}) the Renyi entropies can be computed analogously by replacing $g(s)$ by the function
\be
g_n(s)=\frac{1}{n-1}\, \log\left[(\coth(\pi s)/2+1/2)^n-(\coth(\pi s)/2-1/2)^n\right]\,,\label{renre}
\ee
with $g_1(s)=g(s)$. We get the well-known result
\be
S_n=\frac{n+1}{12\, n} \log\left(\frac{b-a}{\epsilon}\right)\,.
\ee

\section{The two interval case} \label{dosintervalos}

To start, we need first to know the expression of the kernel $C^{-1}\equiv\delta^{-1}(x-y)$ for two intervals. The commutator is block diagonal in each of the intervals, and we get the same result as (\ref{dit}) for each of the intervals separately,
\be
(\delta')^{-1}(x,y)= \left\{ \begin{array}{c}
\frac{1}{2}\left(\theta(x-y)-\theta(y-x)\right) \hspace{.7cm} \textrm{if} \,\, x,y \in (a_1,b_1)  \\
\frac{1}{2}\left(\theta(x-y)-\theta(y-x)\right) \hspace{.7cm} \textrm{if} \,\, x,y \in (a_2,b_2)   \\
0 \hspace{.7cm} \textrm{if} \,\, x\in(a_1,b_1)\,, y\in (a_2,b_2)\,\, \textrm{or}\,\, y\in(a_1,b_1)\,, x\in (a_2,b_2)
\end{array} \right.\,,
\ee
or equivalently,
\be
C^{-1}\equiv (\delta')^{-1}(x,y)=\frac{1}{2}\left(\theta(x-y)-\theta(y-x)\right)-\frac{1}{2} \theta(x-a_2)\theta(b_1-y)+\frac{1}{2} \theta(y-a_2)\theta(b_1-x)\,.\label{dosdos}
\ee
Notice this last equation manages to be antisymmetric and its derivative is the delta function.   

Then we have
\be
2 \pi C^{-1}F\equiv \frac{1}{(x-y-i 0^+ )}+\frac{1}{2} \left(\Theta_1(x) \left( \frac{1}{y-a_1}+\frac{1}{y-b_1}\right)+\Theta_2(x)\left( \frac{1}{y-a_2}+\frac{1}{y-b_2}\right)   \right)\,,\label{u1w}
\ee
and
\be
2 \pi F C^{-1}\equiv \frac{1}{(x-y-i 0^+ )}-\frac{1}{2} \left(\Theta_1(y) \left( \frac{1}{x-a_1}+\frac{1}{x-b_1}\right)+\Theta_2(y)\left( \frac{1}{x-a_2}+\frac{1}{x-b_2}\right)   \right)\,,\label{v1w}
\ee
where $\Theta_1(x)$ and $\Theta_2(x)$ are the characteristic functions of the two intervals, that is, functions with value $1$ inside the first interval (respectively second interval) and zero elsewhere.

We have to deal now with kernels that contain theta functions and it might seem at first glance that the analytic method used in previous sections is not applicable here. However, we will show how to bypass this issue. 

To begin with, let us consider eigenvectors $v_s(x)$ of (\ref{v1w}) satisfying the extra property
\be
\fint_{A_1} dx\, v_s(x)=\fint_{A_2}dx\, v_s(x)=0\,.   \label{cope}
\ee
For such particular eigenvectors the second and third term in (\ref{v1w}) vanishes. At the end we will show this is true in the general case. Under this assumption we have that $v_s$ is an eigenfunction of $(x-y-i 0^+ )^{-1}$. Then we use the same ideas as for a single interval, trying to obtain $v_s$ as a boundary value of an analytic function. We again look for analytic functions $S(z)$ on the complex plane with multiplicative boundary conditions on the two intervals $A$ as in (\ref{disc1}). 
The class of eigenfunctions  $u_s$ of the problem must behave near the end points of the intervals as in the case of a single interval (or the case of the half line). That is, they should behave as pure phase factors of the form $u_s\sim e^{-i s \log(x-a_i)}$ or $u_s\sim e^{i s \log(b_i-x)}$ near the end-points. Their derivative, the $v_s$ functions, should at most have single poles (together with a phase factor) at the end-points. Under this condition the general solution is of the form
\be
S(z)\propto  e^{-i s \tilde{\omega}(z)}\left(\frac{\alpha_1}{z-a_1}+\frac{\alpha_2}{z-b_1}+\frac{\alpha_3}{z-a_2}+\frac{\alpha_4}{z-b_2}\right)\label{form}\,,
\ee
with
\be
\tilde{\omega}(z)=\sum_{i=1}^2 \log\left(\frac{z-a_i}{z-b_i}\right)\,.
\ee
Integrating $S(z)$ along a contours encircling the two intervals and a large circle at infinity, it is not difficult to see  that the integral at infinity is equal to the one over the two intervals, which vanish because of (\ref{cope}). Then this functions must fall as $|z|^{-2}$ to cope with (\ref{cope}), and impose the condition
\be
\alpha_1+\alpha_2+\alpha_3+\alpha_4=0\,.\label{sumacero}
\ee  
Calling $\vec{q}=(a_1,b_1,a_2,b_2)$, we have from (\ref{cope}) the coefficients $\alpha_i$ satisfy in addition
\bea
\sum_1^4  \alpha_i I^1_{q_i}=0\,,\label{sumaaa}\\
\sum_1^4  \alpha_i I^2_{q_i}=0\,,\label{rest}
\eea
where
\be
I^l_{q_i}=\fint_{a_l}^{b_l} dx\, e^{-i s \omega(x)}\frac{1}{x-q_i}\,,\hspace{.7cm}l=1,2\,,\hspace{.7cm}i=1,2,3,4\,,  \label{inti}
\ee
and 
\be
\omega(x)= \log\left(-\frac{(x-a_1)(x-a_2)}{(x-b_1)(x-b_2)}\right)\,.
\ee

Only two of the eqs. (\ref{sumacero}), (\ref{sumaaa}) and (\ref{rest}) are independent. This follows from the fact that 
\be
\oint dz\, e^{-i s \omega(z)}\left(\frac{1}{z-q_i}-\frac{1}{z-q_j}\right)=I^1_{q_i}-I^1_{q_j}+I^2_{q_i}-I^2_{q_j}=0\,.\label{fact}
\ee
This complex integral around the two cuts is zero because it is equal to the integral at infinity which vanishes because the integrand falls fast enough.

Therefore we conclude the dimension of the space of these solutions for fixed $s$ is two. The same argument of the previous section shows that these solutions give the eigenvectors of $V^\dagger$ once evaluated on $A$, and any eigenvector of $V^\dagger$ with at most simple poles at the end of the intervals, and satisfying (\ref{cope}), are of this form. 

Now a simple solution is
\be
\tilde{v}_1(z)\propto (e^{-i s \tilde{\omega}})'= -i \,s \,  e^{-i s \tilde{\omega}} \left(\frac{1}{z-a_1}-\frac{1}{z-b_1}+\frac{1}{z-a_2}-\frac{1}{z-b_2}\right)\,.
\ee
In fact, this satisfies (\ref{sumacero}) and (\ref{sumaaa}) because it is proportional to a derivative of the phase $e^{-i s \tilde{\omega}}$ and hence the integral on any of the intervals vanish with the regularization we are using. That is, integrating this function on the intervals we have a further relation for the $I^l_{q}$ integrals,
\bea
I^1_{a_1}-I^1_{b_1}+I^1_{a_2}-I^1_{b_2}=0\,,\label{dely}\\
 I^2_{a_1}-I^2_{b_1}+I^2_{a_2}-I^2_{b_2}=0\,.\label{delyout}
\eea 

The eigenvector $v_1(x)$ follows from taking the boundary limit of $\tilde{v}_1(z)$ on $A$. The corresponding $u_1$ solution is an integral of this function,
\be
u_1= -i\, \textrm{sign} (s) \,C^{-1}\, v_1\propto \, e^{-i s \omega}\,, \label{reluv}
\ee
where in applying (\ref{dosdos}) to $v_1$ boundary terms that are oscillatory phases are discarded, in accordance with the regularization we are using. We can check more directly this is an eigenfunction of (\ref{u1w}) by noting that  
\be
\fint_A dy\, \left( \frac{1}{y-a_1}+\frac{1}{y-b_1}\right)\, u_1(y)=\fint_A dy\, \left( \frac{1}{y-a_2}+\frac{1}{y-b_2}\right)\, u_1(y)\,.
\ee
This follows from (\ref{fact}), (\ref{dely}) and (\ref{delyout}). Hence, the two terms with characteristic functions in (\ref{u1w}) are equal and we can eliminate these functions altogether by replacing $\Theta_1(x),\Theta_2(x)\rightarrow 1/2$. After this replacement the proof that $u_1$  is an eigenvector of the kernel  (\ref{u1w}) follows the same steps as the one for a single interval in the previous section by promoting $u_1$ to $\tilde{u}_1(z)\propto e^{-i s \tilde{\omega}}$ in the complex plane.

We choose the second solution, $\tilde{v}_2(z)$, of the form (\ref{form}), such that its value on $A$ gives a $v_2$ eigenfunction orthogonal to $u_1$. Collecting the coefficients of the would-be delta functions in the scalar product between $u_1$ and $v_2$, which are generated by the integral near the endpoints of the intervals, we have for $\tilde{v}_2$   
\be
\alpha_1-\alpha_2+\alpha_3-\alpha_4=0\,.
\ee
From this and (\ref{sumacero}) we get
\be
\tilde{v}_2\propto e^{-i s w(z)}\left(\frac{1}{z-a_1}+\frac{\alpha}{z-b_1}-\frac{1}{z-a_2}-\frac{\alpha}{z-b_2}\right)\,.
\ee
and from (\ref{sumaaa}) it follows
\be
\alpha = -\frac{I^1_{a_1}-I^{1}_{a_2}}{I^1_{b_1}-I^1_{b_2}}  \,.\label{espre}
\ee

In order to compute $u_2$ we use the fact that the integrals of $v_2$ along the two intervals $A_1$ and $A_2$ vanish according to (\ref{cope}). 
Therefore using $u=-i \textrm{sign}(s)\, C^{-1} v$, and (\ref{v1w}), we have 
\be
u_2(x)=-i \, \textrm{sign}(s) \,\left\{\begin{array}{c} \fint_{a_1}^x dy\, v_2(y)=\fint_{b_1}^x dy\, v_2(y)\,,  \hspace{.7cm} x\in A_1\\ \fint_{a_2}^x dy\, v_2(y)=\fint_{b_2}^x dy\, v_2(y)\,,
                               \hspace{.7cm} x\in A_2
\end{array}\right. \,.
\ee

In order to normalize the solutions, we compute the coefficient of the delta function in the scalar product, that can only come from the singular part of the integrals near the end points of the intervals. We have that the normalized solutions satisfying (\ref{cuen}) are
\bea
v_1(x) &=& \sqrt{\frac{|s|}{4 \pi}} \, e^{-i s \omega(x)} \left(\frac{1}{x-a_1}-\frac{1}{x-b_1}+\frac{1}{x-a_2}-\frac{1}{x-b_2}\right)\,, \label{v1}\\ 
u_1(x) &=& \frac{1}{\sqrt{4\pi|s|}} e^{-i s \omega(x)}\,, \label{u1} \\ 
v_2(x) &=& \sqrt{\frac{|s|}{4 \pi}}\, e^{-i s \omega(x)}\left(\frac{1}{x-a_1}+\frac{\alpha}{x-b_1}-\frac{1}{x-a_2}-\frac{\alpha}{x-b_2}\right)\,,\label{v2} \\
u_2(x) &=& 
-i\frac{s}{\sqrt{4\pi|s|}} \,\left\{\begin{array}{c} \fint_{a_1}^x dy\, e^{-i s \omega(y)}\left(\frac{1}{y-a_1}+\frac{\alpha}{y-b_1}-\frac{1}{y-a_2}-\frac{\alpha}{y-b_2}\right)\,, \hspace{.7cm} x\in A_1\\ \fint_{a_2}^x dy\, e^{-i s \omega(y)}\left(\frac{1}{y-a_1}+\frac{\alpha}{y-b_1}-\frac{1}{y-a_2}-\frac{\alpha}{y-b_2}\right)\,, \hspace{.7cm} x\in A_2 \label{u2}
\end{array}\right. \,.
\eea

We have not yet shown that these are the only possible eigenfunctions, since we imposed the additional conditions (\ref{cope}) to derive them. We will show at the end of the next subsection these conditions follow from orthogonality with $u_1^s$ and $u_2^s$ in the limit $|s|\rightarrow 0$. To do that we first have to get simpler expressions for these eigenvectors. 

\subsubsection{Dependence of the eigenvectors through the cross ratio}

The aim of this section is to obtain simplified expressions for the function $\alpha$ \eqref{espre} and the eigenfunctions (\ref{v1}--\ref{u2}), which will be useful for the final computation of the modular Hamiltonian and the mutual information. For such purpose, we analyse the behaviour of such expressions in terms of the cross ratio
\be
\eta=\frac{(b_1-a_1)(b_2-a_2)}{(a_2-a_1)(b_2-b_1)}\in (0,1)\, , \label{crossratio}
\ee
which is the natural geometric parameter of the problem given the conformal invariance of the model. We consider a change of coordinates $x' = f(x)$ given by a general Mobi\"us transformation 
\be
x'=f(x) = \frac{a x +b}{c x +d}\, , \label{mobius}
\ee
where $a,b,c,d \in \mathbb{R}$ and $a d-c b > 0$. Such transformation leaves the cross ratio \eqref{crossratio} invariant.

Let us first understand the dependence of the function $\alpha(s;a_1,b_1,a_2,b_2)$ with the interval endpoints. We can use \eqref{mobius} to make a change in the integration variables on the integrals \eqref{inti} involved on the definition of the function $\alpha$. After that, an straightforward computation shows that 
\be
\alpha(s;a_1,b_1,a_2,b_2)=\alpha(s;a'_1,b'_1,a'_2,b'_2) \,, 
\ee
where the two sets of intervals endpoints are related by
\be
a'_i=f(a_i)\,,\,\,\,\, b'_i=f(b_i) \hspace{1cm} \text{ for } i=1,2 \, .  \label{epr}
\ee
Since such relation holds for any general Mobi\"us transformation, we have that  $\alpha(s,\eta)$ depends on the intervals endpoints only through the cross ratio.

Similarly, a direct computation for the eigenfunctions reveals the following covariance properties under the change of variables $x'=f(x)$,
\bea
u_i(x';q'_i) &=& \mathrm{e}^{i s K(q_i)} \, u_i(x;q_i) \, , \label{uim} \\  
v_i(x'; q'_i ) &=& \mathrm{e}^{i s K(q_i)} \frac{1}{f'(x)} u_i(x; q_i) \label{vim} \, ,   
\eea
where $u_i(x;q_i)$ and $v_i(x;q_i)$ are the eigenfunctions corresponding to the problem with endpoints $q_i = (a_1,b_1,a_2,b_2)$ (idem for $q'_i$) and the two sets of endpoints are related by \eqref{epr}. The real function $K(q_i)$ is given by
\be
K(q_i) = \frac{1}{2} \log \left( \frac{f'(a_1)f'(a_2)}{f'(b_1)f'(b_2)} \right) \, .
\ee
Not surprisingly, the $u$ functions transform as a scalar wave, and the functions $v$ as their derivatives under conformal transformations.

Simpler expressions are obtained when we specially take the Mobi\"us transformation $f_1(x)$ which sends the points $(a_1,b_1,a_2,b_2) \rightarrow (0,\eta, 1, \infty)$, i.e.\footnote{More carefully, we shall take the Mobi\"us transformation which transforms $(a_1,b_1,a_2,b_2) \rightarrow (0,\eta, 1, \Lambda)$ with $\Lambda > 1$ and in the end take $\Lambda \rightarrow \infty$.} 
\bea
f_1(x) &:=& \frac{\left(b_{2}-a_{2}\right)\left(x-a_{1}\right)}{\left(a_{2}-a_{1}\right)\left(b_{2}-x\right)} \, , \label{f1}\\
f'_1(x) &=& \frac{\left(b_{2}-a_{1}\right)\left(b_{2}-a_{2}\right)}{\left(a_{2}-a_{1}\right)\left(b_{2}-x\right)^{2}}=\frac{1}{b_{2}-a_{1}}\left(\frac{b_{2}-a_{2}}{a_{2}-a_{1}}+2x'+\frac{a_{2}-a_{1}}{b_{2}-a_{2}}x'^{2}\right)   \, .
\eea
With this transformation we get from (\ref{espre}) the compact formula
\be 
\alpha(s,\eta) = -\frac{_{2}F_{1}\left(1+is,-is;1;\eta\right)}{_{2}F_{1}\left(1-is,is;1;\eta\right)} \, , \label{alphah}
\ee
where $_{2}F_{1}(a,b;c;x)$ is the Gaussian or ordinary hypergeometric function. To derive this result we used the integral representation for such function,
\be
_{2}F_{1}(a,b;c;x) = \frac{\Gamma(c)}{\Gamma(b)\Gamma(c-b)} \int_0^1 dt \, t^{b-1} (1-t)^{c-b-1} (1- t x)^{-a}  \, , \label{hyper_def}
\ee
for $x<1$ and Re$(c)>$ Re$(b)>0$.\footnote{When Re$(b)=0$, which occurs in \eqref{alphah}, equation \eqref{hyper_def} has to be understood as the $\fint$-regularization explained on \eqref{fint_reg}.  \label{foot-hreg}} Expression \eqref{alphah} reveals the dependence of $\alpha$ through the cross ratio, and the fact that $\alpha$ is a phase factor.  For $s=0$ we have $\alpha=-1$, and it reaches a value dependent on $\eta$ for $s\rightarrow\pm \infty$ (see below). We also have $\alpha(-s,\eta)=\alpha(s,\eta)^*$. For fixed $s\neq 0$ we have $\lim_{\eta\rightarrow 0} \alpha=-1$, and $\lim_{\eta\rightarrow 1} \alpha=1$. 

Applying the same transformation to the eigenvectors (\ref{v1}--\ref{u2}) and using the expressions (\ref{uim}--\ref{vim}) we arrive to\footnote{Here we drop out a global constant phase factor which is the same for all the eigenvectors, without modifying the orthonormalization condition and the condition \eqref{cuen}.}
\bea
u_1(x) &=&  \frac{1}{\sqrt{4\pi|s|}} \mathrm{e}^{-i s \log\left(\frac{x'(1-x')}{\eta-x'}\right)}  \,, \label{u1cr} \\
v_1(x) &=&  \sqrt{\frac{|s|}{4\pi}}\, f'_1(x) \,\mathrm{e}^{-i s \log\left(\frac{x'(1-x')}{\eta-x'}\right)}\left(\frac{1}{x'}+\frac{1}{x'-1}-\frac{1}{x'-\eta}\right) \, , \label{v1cr} \\
u_2(x) &=&  i\frac{s}{\sqrt{4\pi|s|}}\,\left(\frac{x'}{\eta}\right)^{-i s}\Bigg[\frac{1}{i s}F_{1}\left(-i s;i s,-i s;1-i s;x',\frac{x'}{\eta}\right) + 
 \frac{\alpha}{1-is} \, \frac{x'}{\eta} \, F_{1}\left(1-i s;i s,1-i s;2-i s;x',\frac{x'}{\eta}\right) 
\nonumber \\
&& -\frac{x'}{1-i s}F_{1}\left(1-i s;1+i s,-i s;2-i s;x',\frac{x'}{\eta}\right)\Bigg]    \,, \hspace{1 cm } x\in A_1 \, , \label{ua1i} \\
v_2(x) &=& \sqrt{\frac{\left|s\right|}{4\pi}} \,f'_1(x)\, \mathrm{e}^{-i s \log\left(\frac{x'(1-x')}{\eta-x'}\right)}\left(\frac{1}{x'}+\frac{\alpha}{x'-\eta}-\frac{1}{x'-1}\right)\, , \label{v2cr}
\eea
where the function $F_{1}(a;\beta_1,\beta_1;c;z_1,z_2)$ in \eqref{ua1i} is the Appell Hypergeometric function of two variables. Such function has the following integral representation 
\be
F_{1}(a;\beta_1,\beta_1;c;z_1,z_2)= \frac{\Gamma(c)}{\Gamma(a)\Gamma(c-a)}\int_0^1\,t^{a-1}(1-t)^{c-a-1}(1-tz_1)^{-\beta_1}(1-tz_2)^{-\beta_2} \, ,
\ee
for $x,y<1$ and Re$(a)>0$ and Re$(c-a)>0$.\footnote{For Re$(a)=0$, see footnote \ref{foot-hreg}.} We remark formula \eqref{ua1i} is only valid for $x \in A_1$, and in such case $x' \in (0,\eta)$ and hence the arguments of the Appell's function belong to its domain of analyticity. 

To get an expression for $u_2$ valid for $x \in A_2$ in terms of Appell functions we must consider a different Mobi\"us transformation $\tilde{x} :=f_2(x) $ which sends $(a_1,b_1,a_2,b_2) \rightarrow (1,\infty,0,\eta)$,
 \bea
f_2(x) &:=& \frac{(b_1-a_1)(x-a_2)}{(a_2-a_1)(x-b_1)} \, , \label{f2}  \\
f'_2(x) &=& \frac{(b_1-a_1)(a_2-b_1)}{(a_2-a_1)(x-b_1)^2} = \frac{1}{a_2-b_1}\left(\frac{b_1-a_1}{a_2-a_1}+2\tilde{x}+\frac{a_2-a_1}{b_1-a_1}\tilde{x}^2\right)   \, .
\eea
Then we obtain for $x \in A_2$,
\bea
u_2(x) &=&  -i\frac{s}{\sqrt{4\pi|s|}}\,\left(\frac{\tilde{x}}{\eta}\right)^{-i s}\Bigg[\frac{1}{i s}F_{1}\Bigg(-i s;i s,-i s;1-i s;\tilde{x},\frac{\tilde{x}}{\eta}\Bigg) + \label{uiui}\\
&& \frac{\alpha}{1-i s} \, \frac{\tilde{x}}{\eta} \, F_{1}\Bigg(1-i s;i s,1-i s;2-i s;\tilde{x},\frac{\tilde{x}}{\eta}\Bigg) 
-\frac{\tilde{x}}{1-i s}F_{1}\Bigg(1-i s;1+i s,-i s;2-i s;\tilde{x},\frac{\tilde{x}}{\eta}\Bigg)\Bigg]    \, ,    \nonumber
\eea
which is the same expression valid for $u_2$ in the first interval (up to a minus global sign) but evaluated in $\tilde{x}$ instead of $x'$. Such expression indicates that for any point $x_1 \in A_1$ exits a point $x_2 \in A_2$ such $u_2(x_1)= - u_2(x_2)$, and viceversa. In the next subsection we show that all the eigenvectors are classified according to such  ``parity symmetry''.

\subsubsection{Parity symmetries of the eigenfunctions} \label{psym}
\label{parity}

In this section we study the behaviour of the eigenfunctions under a conformal transformation that interchanges the two intervals. For that we introduce the Mobi\"us transformation 
\be 
\bar{x}=p(x) \label{reflec}
\ee
 which interchange the two intervals $(a_1,b_1,a_2,b_2) \leftrightarrow (a_2,b_2,a_1,b_1)$, namely
\bea
p(x) &=& \frac{a_1a_2(x-b_1-b_2)-b_1b_2(x-a_1-a_2)}{x(a_1+a_2-b_1-b_2)+(b_1b_2-a_1a_2)} \,, \label{moebius-i} \\
p'(x) &=& \frac{(b_1-a_1)(b_2-a_1)(a_2-b_1)(a_2-b_2)}{[x(a_1+a_2-b_1-b_2)+(b_1b_2-a_1a_2)]^2}>0   \, ,
\eea
where \eqref{moebius-i} is the same as  \eqref{conj-points}, what indicates that $\bar{x}$ is the conjugate point of the point $x$, that is,  $\omega(\bar{x})=\omega(x)$. Specializing this transformation on relations (\ref{uim}-\ref{vim}) we get
\bea
u_i(\bar{x};a_2,b_2,a_1,b_1) &=&  u_i(x;a_1,b_1,a_2,b_2) \, , \label{uip} \\  
v_i(\bar{x};a_2,b_2,a_1,b_1)) &=& \frac{1}{p'(x)} u_i(x;a_1,b_1,a_2,b_2) \label{vip} \, ,   
\eea
where in this case we have $K(a_1,b_1,a_2,b_2)=0$. On the other hand, from (\ref{v1}-\ref{v2}) we easily see that 
\bea
u_1(x;a_2,b_2,a_1,b_1) &=&  u_1(x;a_1,b_1,a_2,b_2) \, ,  \\
v_1(x;a_2,b_2,a_1,b_1) &=&  v_1(x;a_1,b_1,a_2,b_2) \, ,  \\  
v_2(x;a_2,b_2,a_1,b_1) &=&  -v_2(x;a_1,b_1,a_2,b_2) \,  .   
\eea
Therefore, using additionally (\ref{ua1i}) and (\ref{uiui}) for $u_2$, we conclude we have the following parity symmetries
\bea
u_1(\bar{x}) &=&  u_1(x) \, ,  \label{u1ps} \\  
v_1(\bar{x}) &=&  \frac{1}{p'(x)} v_1(x) \, , \label{v1ps} \\  
u_2(\bar{x}) &=&  - u_2(x) \,  ,   \label{u2ps} \\
v_2(\bar{x}) &=&  - \frac{1}{p'(x)} v_2(x) \,  .   \label{v2ps} 
\eea
Then, the first set of eigenfunctions is even and the second is odd under taking the conjugate point $\bar{x}$. We remark in these expressions we have the same eigenfunctions for the same endpoints $(a_1,b_1,a_2,b_2)$ appear at both sides.

Another particular Mobi\"us transformation which implies a quite simple symmetry relation, is the transformation $\hat{x}=q\left(x\right)$ which sends the end points $\left(a_{1},b_{1},a_{2},b_{2}\right)\rightarrow\left(b_{1},a_{1},b_{2},a_{2}\right)$, i.e. reflects each interval into itself,
\be
\hat{x}=q\left(x\right)=\frac{x(a_{2}b_{2}-a_{1}b_{1})-a_{1}a_{2}(b_{2}-b_{1})-b_{1}b_{2}(a_{2}-a_{1})}{x(b_{2}+a_{2}-b_{1}-a_{1})+a_{1}b_{1}-a_{2}b_{2}}\,. \label{tranfi}
\ee
Under such transformation, the eigenvectors satisfy
\bea
u_{1,-s}\left(\hat{x}\right) & = & \mathrm{e}^{isK\left(q_{i}\right)}u_{1,s}\left(x\right)\,,\label{uuu1}\\
u_{2,-s}\left(\hat{x}\right) & = & \mathrm{e}^{isK\left(q_{i}\right)}\left(-1\right)\alpha\left(-s\right)u_{2,s}\left(x\right)\,,\label{uuu2}\\
v_{1,-s}\left(\hat{x}\right) & = & \mathrm{e}^{isK\left(q_{i}\right)}\frac{\left(-1\right)}{q'\left(x\right)}v_{1,s}\left(x\right)\,,\\
v_{2,-s}\left(\hat{x}\right) & = & \mathrm{e}^{isK\left(q_{i}\right)}\frac{\alpha\left(-s\right)}{q'\left(x\right)}v_{2,s}\left(x\right)\,,
\eea
where now we must explicitly write the dependence of the eigenfunctions with the parameter $s$ because the above expressions relate a eigenfunction of eigenvalue $s$ with the one of eigenvalue $-s$. In this case we have a no null phase factor $K\left(q_{i}\right)=2\log\left(\frac{b_{2}-b_{1}}{a_{2}-a_{1}}\right)$. 

\subsubsection{Completeness of the eigenvector system}

Before we compute the mutual information and modular Hamiltonian, we have to clarify if the eigenvector basis is complete. When we explicitly constructed the eigenvectors, we only consider solutions satisfying the equation \eqref{cope} in order to simplify the calculation, but there was no further reason to assume that. Now, we are be able to show that any other possible eigenvector must satisfy \eqref{cope}, and hence there are no other eigenvectors than the ones already obtained. This fact follows considering the $s=0$ solutions for $u_1$ and $u_2$. Taking out an irrelevant factor of $|s|^{-1/2}$ (which is compensated by the inverse factor in the eigenfunctions $v$) we have
\bea
&& \lim_{s\rightarrow 0} \sqrt{4\pi |s|} \, u_1(x,s)=1\,,   \\
&& \lim_{s\rightarrow 0} \sqrt{4\pi |s|} \, u_2(x,s)= \Theta_1(x)-\Theta_2(x)\,.  
\eea
The first one is proportional to a constant, the same in the two intervals, while the second one is proportional to two opposite constants in the two different intervals. Hence, any third solution $v_3$ for any $s$ would be orthogonal to $u_1(s=0)$ and $u_2(s=0)$, and therefore must satisfy (\ref{cope}). Therefore there cannot be any other eigenvectors for two intervals.

\subsection{Mutual information}

In this section we compute the mutual information using the formulas developed in the previous sections. As we did for the one interval case, the equivalent expression of \eqref{s1int} to the present case is
\be
S (A)  =  \sum_{k=1}^2 \int_{A^{(\epsilon)}} dx\int_{0}^{+\infty}ds\,u_{k,s}(x)v_{k,s}^{*}(x) \, g(s) \, , \label{s2int}
\ee
where $A^{(\epsilon)}=A_1^{(\epsilon)} \cup A_2^{(\epsilon)}$ with $A_i^{(\epsilon)}=(a_i+\epsilon, b_i-\epsilon)$ is the UV-regularized region and $g(s)$ is given by \eqref{gs}. The entropy is UV-divergent, and is more convenient to express the result in terms of the mutual information
\be
I(A_1,A_2) = S(A_1) + S(A_2) - S(A_1 \cup A_2) \, , \label{mi-def}
\ee
which is finite and independent of regularization. The one interval entropies $S(A_i)$ are obtained from \eqref{s1}. Each term involved on equation \eqref{mi-def} is UV-divergent and its corresponding regularizations cannot be chosen independently. They must correspond to evaluate the integrals \eqref{s1int} and \eqref{s2int} along the regularized regions as we have already defined, with the same cutoff parameter $\epsilon$ for all the terms. After that, we take the limit $\epsilon \rightarrow 0^+$ and we get the finite desired result for the mutual information.

\begin{figure}[t]
\begin{center}
\includegraphics[width=0.6\textwidth]{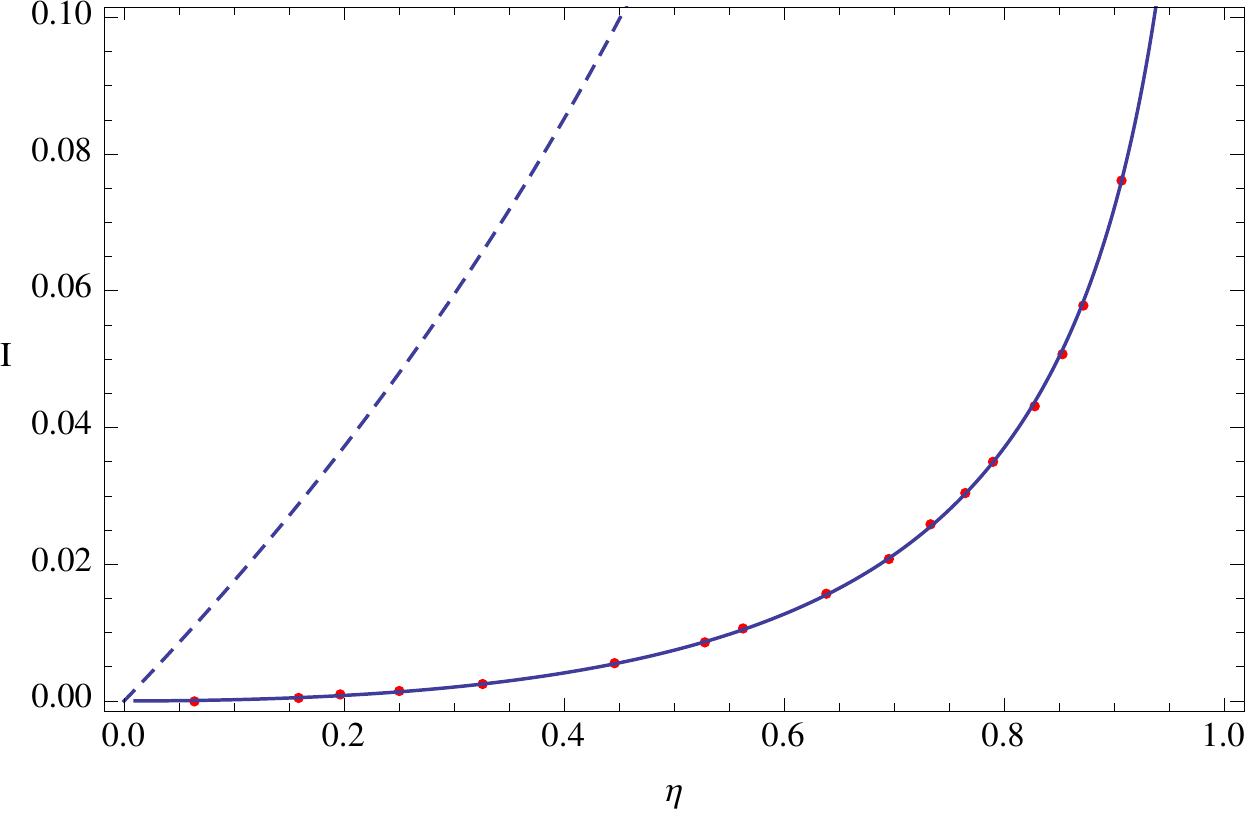}
\captionsetup{width=0.9\textwidth} 
\caption{The mutual information $I(A_1,A_2)$ as function of the cross ratio $\eta$. The continuous solid line corresponds to the mutual information obtained by numerical integration of the analytic expression (\ref{mic}), (\ref{Ueta}). The red points correspond  to the simulation on the lattice as it is explained on section \ref{numerico}. The dashed line is the chiral free fermion mutual information $-(1/6) \log(1-\eta)$.}
\label{figufi}
\end{center}
\end{figure}

Using the formulae (\ref{v1}-\ref{u1}), the first term on \eqref{s2int} can be easily calculated
\bea
S_1(A) & = & \int_{A^{(\epsilon)}} dx\int_{0}^{+\infty}ds\,u_{1,s}(x)v_{1,s}^{*}(x) \, g(s)  =  \frac{1}{24}\int_{A^{\left(\epsilon\right)}}dx\,\omega'(x) \\
 & = &\frac{1}{12}\log\left(\frac{\left(a_{2}-b_{1}\right)\left(b_{2}-a_{1}\right)}{\left(b_{2}-b_{1}\right)\left(a_{2}-a_{1}\right)}\right)+\frac{1}{12}\log\left(\frac{b_{1}-a_{1}}{\epsilon}\right)+\frac{1}{12}\log\left(\frac{b_{2}-a_{2}}{\epsilon}\right) \\
 &=& \frac{1}{12} \log(1-\eta) + \frac{1}{2} S(A_1) + \frac{1}{2} S(A_2) \,. 
\eea

\begin{figure}[t]
\begin{center}
\includegraphics[width=0.6\textwidth]{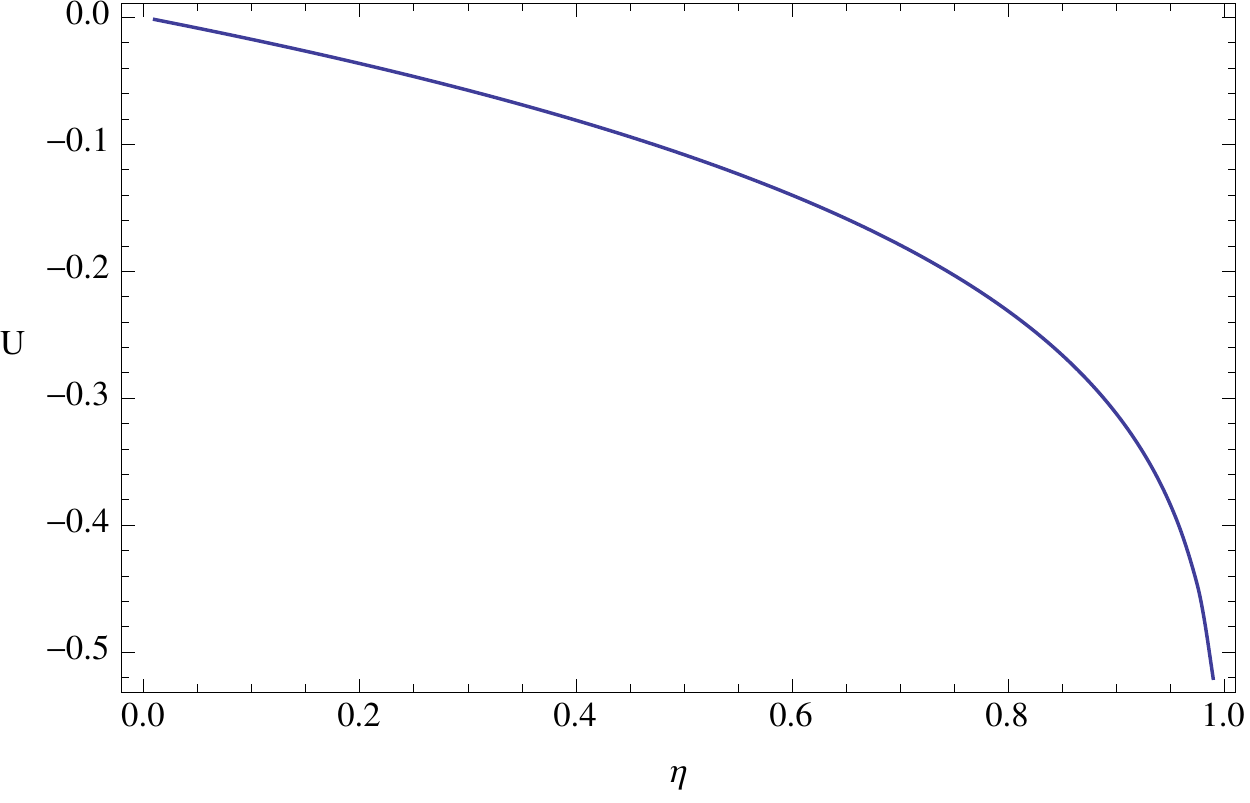} 
\captionsetup{width=0.9\textwidth}
\caption{The function $U(\eta)$ as a function of the cross ratio $\eta$. This function is negative because the 
model of the current is a subalgebra of the free fermion model. It does not have the $\eta\leftrightarrow (1-\eta)$ symmetry expected for the case where the entropy for two intervals is equal to the entropy of its complement. $U(\eta)$ has a $-1/2 \log(-\log(1-\eta))$ divergence for $\eta\rightarrow 1$.}
\label{figuu}
\end{center}
\end{figure}

Direct treatment of the integral for the second eigenvector is more complicated due to the presence of hypergeometric and Appell functions in $u_2$. We will find convenient to use the following trick instead. Since the integral in (\ref{s2int}) is regularized, keeping $\epsilon$ fixed, we can replace 
\bea
&& \int_{A^{(\epsilon)}} dx\,u_{2,s}(x)v_{2,s}^{*}(x)=\lim_{\delta s\rightarrow 0} \int_{A^{(\epsilon)}} dx\,u_{2,s}(x)v_{2,s+\delta s}^{*}(x)\label{tri}\\
&& =\lim_{\delta s\rightarrow 0} \left(\int_{A} dx\,u_{2,s}(x)v_{2,s+\delta s}^{*}(x)
-\sum_{i=1,2}\left(\int_{a_i}^{a_i+\epsilon}dx\, u_{2,s}(x)v_{2,s+\delta s}^{*}(x)+\int_{b_i-\epsilon}^{b_i}dx\,u_{2,s}(x)v_{2,s+\delta s}^{*}(x) \right) \right)\nonumber \\
&&=- \lim_{\delta s\rightarrow 0} 
\sum_{i=1,2}\left(\int_{a_i}^{a_i+\epsilon}dx\, u_{2,s}(x)v_{2,s+\delta s}^{*}(x)+\int_{b_i-\epsilon}^{b_i}dx\,u_{2,s}(x)v_{2,s+\delta s}^{*}(x) \right)\,. \nonumber
\eea
In the last step we have used the fact that vectors $u_s$ and $v_{s+\delta s}$ are orthogonal for $\delta s\neq 0$. The advantage is that we do not need now the precise behaviour of the eigenfunctions along the intervals but only in a small region near the end point of the intervals. Then we can just take the leading terms of $u_s$ and $v_{s+\delta s}$ since all other subleading terms in $\epsilon$ will disappear in the limit $\epsilon\rightarrow 0$. From (\ref{v1}--\ref{u2}) these leading terms are
\bea
\hspace{-1.5cm}&& u_{2,s}\sim \frac{(-1)^{i+1}}{\sqrt{4\pi|s|}}\, e^{-i s \omega}\hspace{.7cm}\textrm{for} \,x\sim a_i\,,\hspace{.7cm} u_{2,s}\sim \frac{(-1)^{i}}{\sqrt{4\pi|s|}}\,\alpha(s,\eta)\, e^{-i s \omega}\hspace{.7cm}\textrm{for}\,x\sim b_i\,, \\
\hspace{-1.5cm}&& v_{2,s}\sim (-1)^{i+1}\sqrt{\frac{|s|}{4\pi}}\, \frac{e^{-i s \omega}}{x-a_i}\hspace{.5cm}\textrm{for} \,x\sim a_i\,,\hspace{.5cm} v_{2,s}\sim (-1)^{i+1}\,\alpha(s,\eta)\,\sqrt{\frac{|s|}{4\pi}}\,  \frac{e^{-i s \omega}}{x-b_i}\hspace{.5cm}\textrm{for}\,x\sim b_i\,.
\eea

Plugging this back in (\ref{tri}) we get
\be
S_2(A)=\frac{1}{12} \log(1-\eta) + \frac{1}{2} S(A_1) + \frac{1}{2} S(A_2)- \int_0^{+\infty} ds\, \frac{g(s)}{2\pi}\, i\, \alpha(s,\eta) \partial_s \alpha^*(s,\eta)\,.\label{tio}
\ee
This implies 
\be
I(A_1,A_2)  =  -\frac{1}{6} \log(1-\eta) + U(\eta) \label{mic} \, ,
\ee
where the first term in \eqref{mic} coincides with the mutual information of the free chiral fermion field \cite{modu,fosco}, and, taking into account that $-i \,\alpha(s,\eta) \partial_s \alpha^*(s,\eta)=i \partial_s \log(\alpha)$, and integrating by parts, the second term is given by  
\be
U(\eta)  = -\frac{i \,\pi}{2}\int_{0}^{+\infty} ds\,\frac{s}{\sinh^2(\pi s)}\,\log\left(\frac{_{2}F_{1}\left(1+is,-is;1;\eta\right)}{_{2}F_{1}\left(1-is,is;1;\eta\right)}\right)                \, . \label{Ueta}
\ee
We could not express this last integral in terms of standard known functions, and it has to be performed numerically. The result for $U(\eta)$ is always negative, as it must be, considering that the chiral current is a subalgebra of the chiral fermion algebra,\footnote{By bosonisation the fermion current has exactly the same $n$-point functions as $j(x)$.} and hence the mutual information has to be smaller (see formula (\ref{mic})). In figure \ref{figufi} we show a plot of the mutual information while the function $U(\eta)$ is shown in figure \ref{figuu}.

\begin{figure}[t]
\begin{center}
\includegraphics[width=0.6\textwidth]{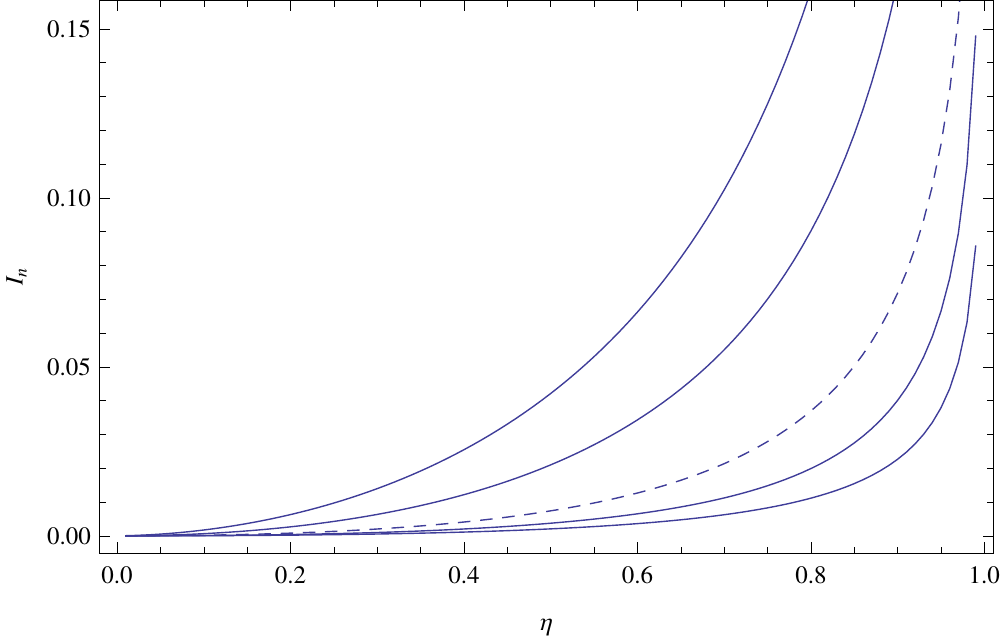} 
\captionsetup{width=0.9\textwidth}
\caption{The Renyi mutual information $I_n(\eta)$ as a function of the cross ratio $\eta$ for different values of $n$. From top to bottom $n=1/3, 1/2, 1, 2, \infty$. The mutual information $I_1(\eta)=I(\eta)$ is shown with the dashed line. }
\label{rer}\end{center}
\end{figure}

The mutual Renyi entropies $I_n(\eta)=S_n(A_1)+S_n(A_2)-S_n(A_1\cup A_2)$ can be computed by replacing $g(s)$ by $g_n(s)$, eq. (\ref{renre}), in (\ref{tio}). Hence we get
\be
I_n(\eta)=-\frac{n+1}{12 \, n}\, \log(1-\eta)+U_n(\eta)\,,
\ee
with
\be
U_n(\eta)=\frac{i\, n}{2(n-1)}\int_{0}^{+\infty} ds\,\left(\coth(\pi s n)-\coth(\pi s)\right)\,\log\left(\frac{_{2}F_{1}\left(1+is,-is;1;\eta\right)}{_{2}F_{1}\left(1-is,is;1;\eta\right)}\right)                \, . \label{Uneta}
\ee
  In figure \ref{rer} we show $I_n(\eta)$ for some values of $n$.

 We will come back to discuss some aspects of these results in the following sections. In the rest of this section we will work out the asymptotic expansions for the mutual information for large and short distances between the intervals.

\subsubsection{Asymptotic behaviour for $I(\eta)$}
 
Before we continue with the calculation of the modular Hamiltonian, we want to analize the asymptotic behaviour of the function $U(\eta)$ in the limits $\eta \rightarrow 0$ and $\eta \rightarrow 1$.

The $\eta\rightarrow 0$ limit corresponds to the large distance limit between the intervals. Since the integrand of (\ref{Ueta}) is analytic at $\eta=0$, a simple Taylor expansion reveals the following asymptotic behaviour
\be
U(\eta)= -\frac{1}{6} \eta - \frac{1}{15} \eta^2 -\frac{13}{315} \eta^3 +\mathcal{O}(\eta^4) \, .
\ee 
This gives
\be
I(\eta)\sim \frac{\eta^2}{60}+ \frac{\eta^3}{70}+\mathcal{O}(\eta^4)\,.
\ee
The first term coicides with the general result for the leading term of the large distance expansion of the mutual information for CFT  \cite{yapeyu1,yapeyu2,cardy5,faulkner}. For two intervals this is
\be
I(\eta)\sim \frac{\sqrt{\pi} \Gamma(2\Delta+1)}{4^{2\Delta+1}\Gamma(2\Delta+3/2)}\eta^{2 \Delta}\,, 
\ee
where $\Delta$ is the lowest dimensional operator of the theory. In the present model this is $j(x)=\partial \phi(x)$ and has $\Delta=1$. The fermion has $\Delta=1/2$ and a different behavior $I(\eta)\sim (1/6)\, \eta$ at large distances, what is quite visible in figure \ref{figufi}.

The short distance limit $\eta \rightarrow 1$ is more tricky, since the integrand of (\ref{Ueta}) converges to zero in a non uniform way in such limit. The main contribution to $U(\eta)$ in this limit comes from  $s\sim 0$.     
 We have to expand the Hypergeometric functions at the numerator and denominator inside the logarithm in (\ref{Ueta})  for $\eta\sim 1$ and $s\sim 0$ to get
\be
U(\eta)\sim -\frac{i \pi}{2}\int_{0}^{+\infty} ds\,\frac{s}{\sinh^2(\pi s)}\,\log\left(\frac{i-s\, \log(1-\eta)}{i+s\, \log(1-\eta)}\right) \sim -\frac{1}{2}\log(-\log(1-\eta)) \,.
\ee
This gives the expansion
\be
I(\eta)\sim -\frac{1}{6}\log(1-\eta)-\frac{1}{2}\log(-\log(1-\eta))\,.
\ee

\subsection{Modular Hamiltonian} \label{sec_mh_2}
We have now all the necessary elements to compute the modular Hamiltonian. 
This is
\be
{\cal H}=\int_A dx\, \int_A dy\, j(x) H(x,y) j(y)\,, 
\ee
were, according to \eqref{mh}, the kernel becomes
\be
H\left(x,y\right) =  \sum_{k=1}^{2}\int_{-\infty}^{\infty}ds\,u_{k,s}\left(x\right)\pi\left|s\right|u_{k,s}^{*}\left(y\right) \, .\label{formass}
\ee 
The kernel is real and symmetric. 

As the expression for this Hamiltonian will turn out to be quite complex, we need to start with some preliminaries about how we are going to express the results. Our first simplification will be to express the kernel in the four sectors $A_1\times A_1$, $A_1\times A_2$, $A_2\times A_1$ and $A_2\times A_2$,  that we call respectively $H_{11}$, $H_{12}$, $H_{21}$, $H_{22}$, in terms of the kernel in the first interval alone, $H_{11}$, using the parity symmetry of the eigenfunctions. 

Let us start by computing the contribution of the first eigenvector $u_1$ to $H_{11}$,  
\be
 \int_{-\infty}^{+\infty}ds\,u_{1,s}\left(x\right)\pi\left|s\right|u_{1,s}^{*}\left(y\right)=\frac{1}{4}\int_{-\infty}^{+\infty}ds\,s\,\mathrm{e}^{is\left(\omega(x)-\omega(y)\right)} 
 =  \frac{\pi}{2}\frac{\delta\left(x-x_{1}\left(\omega\left(y\right)\right)\right)}{\omega'\left(x\right)}
  = \frac{\pi}{2}\omega'(x)^{-1}\,\delta\left(x-y\right) \, ,\label{dell}
\ee
where we set $x,y\in A_1$ and hence we have summed over only one of the roots of $\omega(x)=\omega(y)$ in the delta function. We will find convenient to write 
\be
H_{11}(x,y) = \pi \,\omega'(x)^{-1} \delta(x-y)+  N(x,y) \, ,  \hspace{.8cm}x,y\in A_1\,,\label{toje}
\ee
that is, we have doubled the delta function contribution by $u_1$, and the remaining part is, from (\ref{formass}) and (\ref{dell}),  
\be
N(x,y) = \int_{-\infty}^{+\infty}ds\,\pi |s| \left[u_2(x)u_2^*(y)-u_1(x)u_1^*(y)\right]\,. 
\ee
It will turn out that $N(x,y)$ is a regular distribution.\footnote{It is a locally integrable function.} Hence it gives the purely non local contribution to the modular Hamiltonian.

Now,  we recall the parity symmetry of the eigenfunctions studied in section \ref{parity} under the conformal transformation $\bar{x}=p(x)$ that interchanges the intervals, eq. (\ref{moebius-i}). We have $u_1(\bar{x})=u_1(x)$ and $u_2(\bar{x})=-u_2(x)$. These relations give the following relations between the kernel of the modular Hamiltonian in the different sectors
\bea
H_{12}(x,y) &=& H_{21}(y,x)= -N(x,\bar{y})\,,\hspace{.8cm} x\in A_1\,, \,\,\,y\in A_2\,, \\
H_{22}(x,y) &=& H_{11} (\bar{x},\bar{y})\,,\hspace{2.4cm} x\in A_2\,, \,\,\,y\in A_2\,.\label{tiro}
\eea
These relations, together with (\ref{toje}), reduce the problem to the one of finding the form of the kernel $N(x,y)$ in $A_1\times A_1$.
Unless otherwise stated, in the following we will assume $x$ and $y$ belong to the first interval.

A different parity symmetry, (\ref{tranfi}), implies the kernel $N(x,y)$, $x,y\in A_1$, satisfies 
\be
N(x,y)=N(\hat{x},\hat{y})\,,
\ee
where the transformation $x\rightarrow \hat{x}$ is given by (\ref{tranfi}).  This follows from the corresponding symmetry of the eigenvectors (\ref{uuu1}), (\ref{uuu2}). 

Another simplification is that we can relate all two interval cases with cross ratio $\eta$ between the four end-points of the intervals to the case where the two interval region is the standard region $A_\eta=(0,\eta)\cup (1,\infty)$. This is done using the action of the conformal transformation $x'=f_1(x)$ given by (\ref{mobius}) on the eigenvectors, eq. (\ref{uim}). This simply gives 
\be
N_A(x,y)=N_{\eta}(x',y')\,, \label{right}
\ee
where we wrote explicitly the dependence on the two interval regions. In the following we will call simply $N(x',y')$ to the kernel on the right hand side of (\ref{right}), and keep $x',y'\in (0,\eta)$. That is, we focus on the first interval in the case where the region is $A_\eta$.

An evaluation of $N(x',y')$ requires the integration over $s$, which turns out to be the Fourier transform of products of Appell functions contained in $u_2$, eq.  (\ref{ua1i}).  This obscures the structure of the kernel due to the complexity of these functions, and in particular the analysis of the possible singular terms. Instead we proceed in the following way. We first write the vectors $u_s$ as 
\be
u_s(x')=-i\, \textrm{sign}(s)\,\fint_0^{x'}d\tilde{x}\, v_s(\tilde{x})\,. 
\ee
Then we make the integral in $s$ that will be more amenable. Therefore we write
\be
N(x',y')=\int_0^{x'}d\tilde{x}\, \int_0^{y'}d\tilde{y}\, K(\tilde{x},\tilde{y})\,,\label{ene}
\ee
with
\be
K(\tilde{x},\tilde{y})=\int_{-\infty}^{+\infty}ds\,\pi |s| \left[v_2(\tilde{x})v_2^*(\tilde{y})-v_1(\tilde{x})v_1^*(\tilde{y})\right]\,.\label{masas}
\ee

We split the above kernel in two contributions $K_{1}$ and $K_{2}$  corresponding to $v_1$ and $v_2$. Using \eqref{v1cr} we get
\begin{equation}
K_{1}\left(x',y'\right)  = 
    \frac{\pi}{2}\tilde{\omega}'\left(x'\right)\tilde{\omega}'\left(y'\right)\delta''\left(\tilde{\omega}\left(x'\right)-\tilde{\omega}\left(y'\right)\right)\,,\label{chio0}
\end{equation}
where
\be
\tilde{\omega}\left(x'\right)=\log\left(\frac{x'\left(1-x'\right)}{\eta-x'}\right)\,.
\ee
For the other term we use \eqref{v2cr} and hence
\begin{eqnarray}
\hspace{-1cm} K_{2}\left(x',y'\right) & = & \int_{-\infty}^{\infty}ds\,\pi\left|s\right|\frac{1}{f_{1}'\left(x\right)}\frac{1}{f_{1}'\left(x\right)}v_{2,s}\left(x\right)v_{2,s}^{*}\left(y\right)\nonumber\\
 & = & \frac{1}{4}\int_{-\infty}^{\infty}ds\,s^{2}\,\mathrm{e}^{-is\left(\tilde{\omega}\left(x'\right)-\tilde{\omega}\left(y'\right)\right)}\left(\frac{1}{x'}+\frac{\alpha\left(s,\eta\right)}{x'-\eta}-\frac{1}{x'-1}\right)\left(\frac{1}{y'}+\frac{\alpha\left(-s,\eta\right)}{y'-\eta}-\frac{1}{y'-1}\right)\nonumber \\
 & = & K_{2,\slash\hspace{-.13cm}\alpha}\left(x',y'\right)+K_{2,\alpha}\left(x',y'\right)\,,
\end{eqnarray}
where
\begin{eqnarray}
K_{2,\slash\hspace{-.13cm}\alpha}\left(x',y'\right) & = & -\frac{\pi}{2}\left[\left(\frac{1}{x'}-\frac{1}{x'-1}\right)\left(\frac{1}{y'}-\frac{1}{y'-1}\right)+\frac{1}{x'-\eta}\frac{1}{y'-\eta}\right]\delta''\left(z\right)\,,\label{chio}\\
K_{2,\alpha}\left(x',y'\right) & = & \frac{\pi}{2}\left(\frac{1}{x'}-\frac{1}{x'-1}\right)\frac{1}{y'-\eta}\hat{\alpha}\left(z\right)+\frac{\pi}{2}\frac{1}{x'-\eta}\left(\frac{1}{y'}-\frac{1}{y'-1}\right)\hat{\alpha}\left(-z\right)\,.\label{propi}
\end{eqnarray}
where $z=\tilde{\omega}\left(x'\right)-\tilde{\omega}\left(y'\right)$, and and we have also introduced the function
\begin{eqnarray*}
\hat{\alpha}\left(z,\eta\right) & = & \frac{\text{1}}{2\pi}\int_{-\infty}^{\infty}ds\,s^{2}\alpha\left(s,\eta\right)\,\mathrm{e}^{isz}\,.
\end{eqnarray*}
(\ref{chio}) and (\ref{propi}) are respectively the $\alpha$-independent and $\alpha$-dependent contributions to the kernel $K_{2}\left(x',y'\right)$.

This gives the kernel $N(x',y')$ as a double integral over the sum of (\ref{chio0}), (\ref{chio}) and (\ref{propi}). The final result depends on the Fourier transform of the function $s^{2}\,\alpha\left(s,\eta\right)$, which has to be computed numerically. This numerical computation can be done after we have extracted the leading terms for $s\rightarrow \infty$ from $\alpha(s,\eta)$. This will also help understanding the structure of singularities of these kernels. In the following we will make a further analysis of their local and non local parts. 

\subsubsection{Structure of singular terms}
\begin{figure}[t]
\begin{center}
\includegraphics[width=0.6\textwidth]{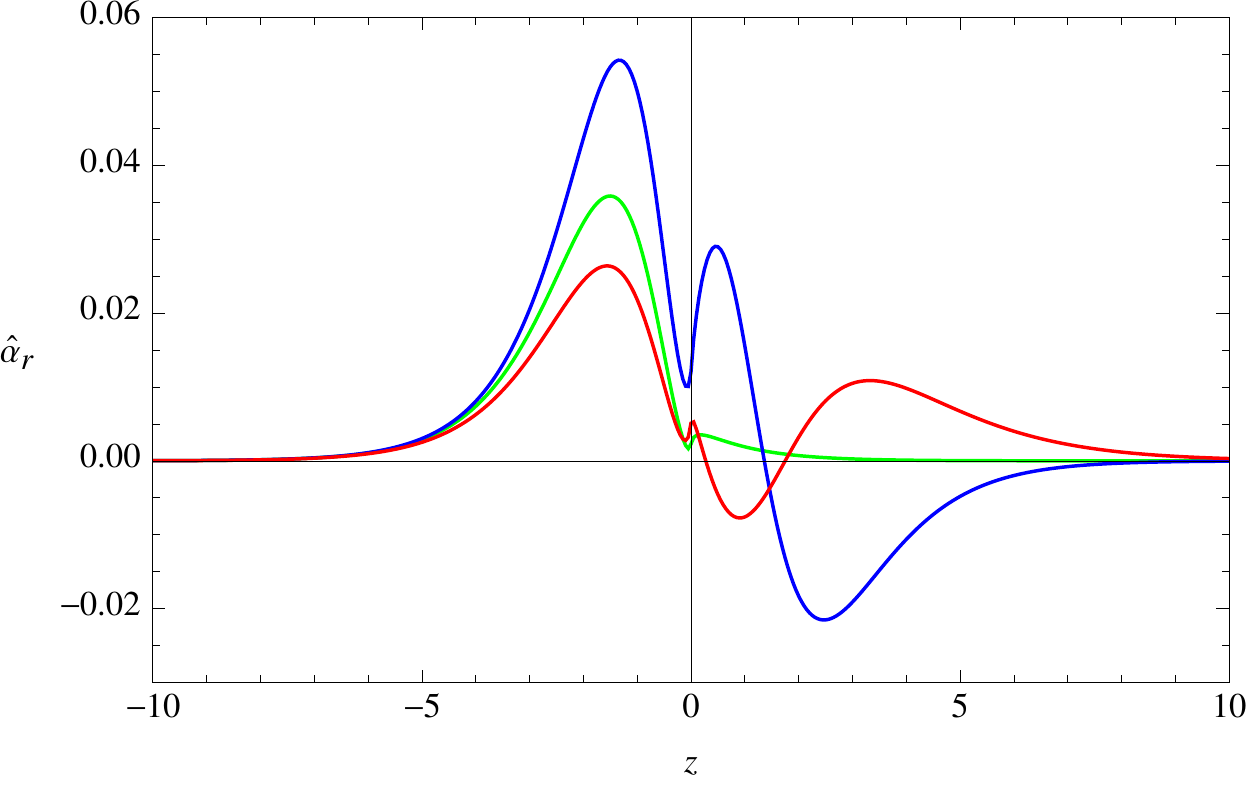} 
\captionsetup{width=0.9\textwidth}
\caption{The function $\hat{\alpha}_r(z)$ for different values of $\eta$, blue $\eta=1/3$, green $\eta=1/2$, and red $\eta=2/3$.}
\label{fourier-alfa}
\end{center}
\end{figure}

The asymptotic behaviour of the hypergeometric functions in $\alpha(s,\eta)$ for large $s$ can be computed using the integral representation and the saddle point approximation. This is straightforward. The leading term was computed for example in \cite{eiko}. Extending this calculation to include fluctuations around the saddle point we get the asymptotic expansion 
\be
\alpha\left(s,\eta\right)=\alpha_{0}+\frac{\alpha_1}{s}+\frac{\alpha_2}{s^{2}}+\frac{\alpha_3}{s^3}+{\cal O}(|s|^{-4})\,, \hspace{.7cm}  |s|\rightarrow \infty\,, \label{falfa}
\ee
where 
\bea
\alpha_{0} & = & \left(2\eta-1\right)+i\,2\sqrt{\eta\left(1-\eta\right)}\,\mathrm{sign}\left(s\right)\,, \\
\alpha_{1} & = & \frac{i}{2}\left(2\eta-1\right)-\sqrt{\eta\left(1-\eta\right)}\, \mathrm{sign}\left(s\right)\,,\\
\alpha_{2} & = & -\frac{i}{16}\frac{1}{\sqrt{\eta\left(1-\eta\right)}}\,\mathrm{sign}\left(s\right)\,, \\
\alpha_3 & = & -\frac{1}{32}\frac{1}{\sqrt{\eta(1-\eta)}}\mathrm{sign} (s)+i\frac{1}{32}\frac{(2\eta-1)}{\eta(1-\eta)}\,.
\eea
Instead of extracting these asymptotic terms directly, we will write 
\be
\alpha\left(s,\eta\right)=\tilde{\alpha}_{0}+\frac{\tilde{\alpha}_{1}}{s}+\frac{\tilde{\alpha}_{2}}{s^{2}}+\frac{\tilde{\alpha}_{3}}{s^{3}}+\alpha_{r}\left(s,\eta\right)\,,
\ee
where now
\bea
\tilde{\alpha}_{0} & = & \left(2\eta-1\right)+i2\sqrt{\eta\left(1-\eta\right)}\,\tanh\left(\frac{\pi}{2}s\right)\,,\\
\tilde{\alpha}_{1} & = & \frac{i}{2}\left(2\eta-1\right)-\sqrt{\eta\left(1-\eta\right)}\,\tanh\left(\frac{\pi}{2}s\right)\,,\\
\tilde{\alpha}_{2} & = & -\frac{i}{16}\frac{1}{\sqrt{\eta\left(1-\eta\right)}}\tanh\left(\frac{\pi}{2}s\right)\,,\\
\tilde{\alpha}_{3} & = & -\frac{1}{32}\frac{1}{\sqrt{\eta(1-\eta)}}\tanh\left(\frac{\pi s}{2}\right)+i\frac{1}{32}\frac{(2\eta-1)}{\eta(1-\eta)}\frac{s^{2}}{s^{2}+1}\,,
\eea
and the reminder function $\alpha_{r}\left(s,\eta\right)$ is smooth in the parameter $s$ and $\alpha_{r}\left(s,\eta\right)\sim\frac{1}{s^{4}}$ when $|s|\rightarrow\infty$. The Fourier transform of $s^2 \alpha(s,\eta)$ is
\bea
\hat{\alpha}\left(z,\eta\right) & = & \frac{1}{2\pi}\int_{-\infty}^{\infty}ds\,s^{2}\alpha\left(s,\eta\right)\,\mathrm{e}^{isz}=\frac{1}{2\pi}\int_{-\infty}^{\infty}ds\,s^{2}\left[\tilde{\alpha}_{0}+\frac{\tilde{\alpha}_{1}}{s}+\frac{\tilde{\alpha}_{2}}{s^{2}}+\frac{\tilde{\alpha}_{3}}{s^{3}}+\alpha_{r}\left(s,\eta\right)\right]\mathrm{e}^{isz}\nonumber\\
 & = & \left(1-2\eta\right)\delta''\left(z\right)+\frac{1}{2}\left(2\eta-1\right)\delta'\left(z\right) +\frac{1}{\pi}\sqrt{\eta\left(1-\eta\right)}\left(3+\cosh\left(2z\right)\right)\text{csch}^{3}\left(z\right)\nonumber\\
 &  & +\frac{1}{2\pi}\sqrt{\eta\left(1-\eta\right)}\sinh\left(2z\right)\,\text{csch}^{3}\left(z\right)+\frac{1}{16\pi}\frac{1}{\sqrt{\eta\left(1-\eta\right)}}\text{csch}\left(z\right)\nonumber\\
 &  & +\frac{1}{32\pi}\frac{1}{\sqrt{\eta(1-\eta)}}\log\left(\tanh\left|\frac{z}{2}\right|\right)+\frac{1}{64}\frac{1-2\eta}{\eta(1-\eta)}\mathrm{sign}\left(z\right)\mathrm{e}^{-\left|z\right|}+\hat{\alpha}_{r}\left(z,\eta\right)\,,
\eea
where $\hat{\alpha}_{r}\left(z,\eta\right)$ is the Fourier transform of $s^{2}\alpha_{r}\left(s,\eta\right)$.  This is a continuous function vanishing exponentially fast at infinity. This real function is computed numerically, and ii is shown in figure \eqref{fourier-alfa} for some values of $\eta$. Putting all together we finally get
\bea
K\left(x',y'\right) & = & K_{1}\left(x',y'\right)+K_{2}\left(x',y'\right)=K_{1}\left(x',y'\right)+K_{2,\alpha}\left(x',y'\right)+K_{2,\slash\hspace{-.13cm}\alpha}\left(x',y'\right)\\
 & = &  k_{\delta''}\left(x',y';\eta\right)\delta''\left(z\right)+\,k_{\delta'}\left(x',y';\eta\right)\delta'\left(z\right)+k_{i}\left(x',y';\eta\right)+k_{r}\left(x',y';\eta\right)\,, \label{Kxys}\nonumber
\eea
where 
\bea
k_{\delta''}\left(x',y';\eta\right) & = & \frac{\pi}{2}\tilde{\omega}'\left(x'\right)\tilde{\omega}'\left(y'\right)-\frac{\pi}{2}\left[\frac{1}{x'\left(1-x'\right)y'\left(1-y'\right)}+\frac{1}{\left(\eta-x'\right)\left(\eta-y'\right)}\right] \nonumber \\
& & +\frac{\pi}{2}\left(2\eta-1\right)\left[\frac{1}{r\left(x',y'\right)}+\frac{1}{r\left(y',x'\right)}\right]\,,\\
k_{\delta'}\left(x',y';\eta\right) & = & -\frac{\pi}{4}\left(2\eta-1\right)\left[\frac{1}{r\left(x',y'\right)}-\frac{1}{r\left(y',x'\right)}\right]\,,\\
k_{i}\left(x',y';\eta\right) & = & \sqrt{\eta(1-\eta)}\frac{\left[r\left(x',y'\right)-r\left(y',x'\right)\right]^{2}}{\left[r\left(x',y'\right)+r\left(y',x'\right)\right]^{3}}+2\sqrt{\eta(1-\eta)}\frac{r\left(x',y'\right)r\left(y',x'\right)}{\left[r\left(x',y'\right)-r\left(y',x'\right)\right]^{2}\left[r\left(x',y'\right)+r\left(y',x'\right)\right]} \nonumber \\
 &  & +\frac{1}{16}\frac{1}{\sqrt{\eta(1-\eta)}}\frac{1}{r\left(x',y'\right)+r\left(y',x'\right)}  -\frac{1}{64}\frac{1}{\sqrt{\eta(1-\eta)}}\frac{r\left(x',y'\right)+r\left(y',x'\right)}{r\left(x',y'\right)r\left(y',x'\right)}\log\left|\frac{r\left(x',y'\right)-r\left(y',x'\right)}{r\left(x',y'\right)+r\left(y',x'\right)}\right| \nonumber \\
 &  & -\frac{\pi}{128}\frac{(2\eta-1)}{\eta(1-\eta)}\left[r\left(x',y'\right)-r\left(y',x'\right)\right]\left[\Theta\left(x-y\right)\frac{1}{r\left(x',y'\right)^{2}}-\Theta\left(y-x\right)\frac{1}{r\left(y',x'\right)^{2}}\right]\label{kirr}\\
k_{r}\left(x',y';\eta\right) & = & -\frac{\pi}{2}\frac{1}{r\left(x',y'\right)}\hat{\alpha}_{r}\left(z,\eta\right)-\frac{\pi}{2}\frac{1}{r\left(y',x'\right)}\hat{\alpha}_{r}\left(-z,\eta\right)\,,\label{kreg}
\eea
where we have defined the positive polynomial function
\be
r\left(x',y'\right)  =  x'\left(1-x'\right)\left(\eta-y'\right)>0\,.
\ee

 The singular terms can be simplified rewriting the Dirac delta distributions in terms of the variable
$x-y$. A careful computation reveals the relations\footnote{The reason for a term containing two derivatives of the
Dirac delta function becomes just a term proportional to $\delta\left(x'-y'\right)$
is that the prefactor function behaves as $k_{\delta''}\left(x',y';\eta\right)\sim\mathcal{O}\left(x'-y'\right)^{2}$
for $x'\rightarrow y'$.}
\bea
\,k_{\delta''}\left(x',y';\eta\right)\delta''\left(\tilde{\omega}\left(x'\right)-\tilde{\omega}\left(y'\right)\right) & = & 2\pi\,\eta \left(1-\eta\right)\frac{x'\left(1-x'\right)\left(\eta-x'\right)}{\left(\eta+x'^{2}-2\eta x'\right)^{3}}\delta\left(x'-y'\right)\,,\label{deltauno}\\
\,k_{\delta'}\left(x',y';\eta\right)\delta'\left(\tilde{\omega}\left(x'\right)-\tilde{\omega}\left(y'\right)\right) & = & \,\frac{\pi}{4}\frac{\left(1-2\eta\right)}{\eta+x'^{2}-2\eta x'}\delta\left(x'-y'\right)\,.\label{deltados}
\eea
Therefore, upon integration in (\ref{ene}) these terms behave as ordinary functions, and they produce a singularity in $N(x',y')$ which is just a jump in the first derivative for $x'=y'$. 

In a similar way, expanding for $x'\sim y'$ the term (\ref{kirr})  
 reveals that it behaves as $(x'-y')^{-2}$  for $x'\rightarrow y'$. Upon integration, this gives a $\log|x'-y'|$ singularity for $N(x',y')$. Hence, this shows that the non local kernel $N(x',y')$ is given by a regular distribution.   

The behaviour of the kernel $K(x',y')$ on the extremes of the interval, for example for $x'\rightarrow 0$ and $y'$ fixed, can be seen more clearly from (\ref{chio0}), (\ref{chio}) and (\ref{propi}). The only non local contribution comes from the Fourier transform of $s^2 \alpha(\eta,s)$, 
\be
K(x',y')\sim \frac{\pi}{2\,x'}\frac{1}{y'-\eta} \hat{\alpha}(\log(x'),\eta)\,,\hspace{.7cm} x'\sim 0\,.
\ee
Since $\alpha(\eta,s)$ is an infinite differentiable function of $s$ its Fourier transform fall faster than any power of the variable. Therefore $K(x',y')$ is integrable on the boundary, and that is the reason we have not needed to use a regularization in (\ref{ene}).\footnote{That is, the regularizing terms in the integral of $v_s$ as in (\ref{fint_reg}) do not contribute if we make the integral in $s$ first.} As a result $N(x',y')$ fall to zero faster than any power of $\log(x')^{-1}$ for $x'\sim 0$.

\begin{figure}[t]
\begin{center}
\includegraphics[width=0.6\textwidth]{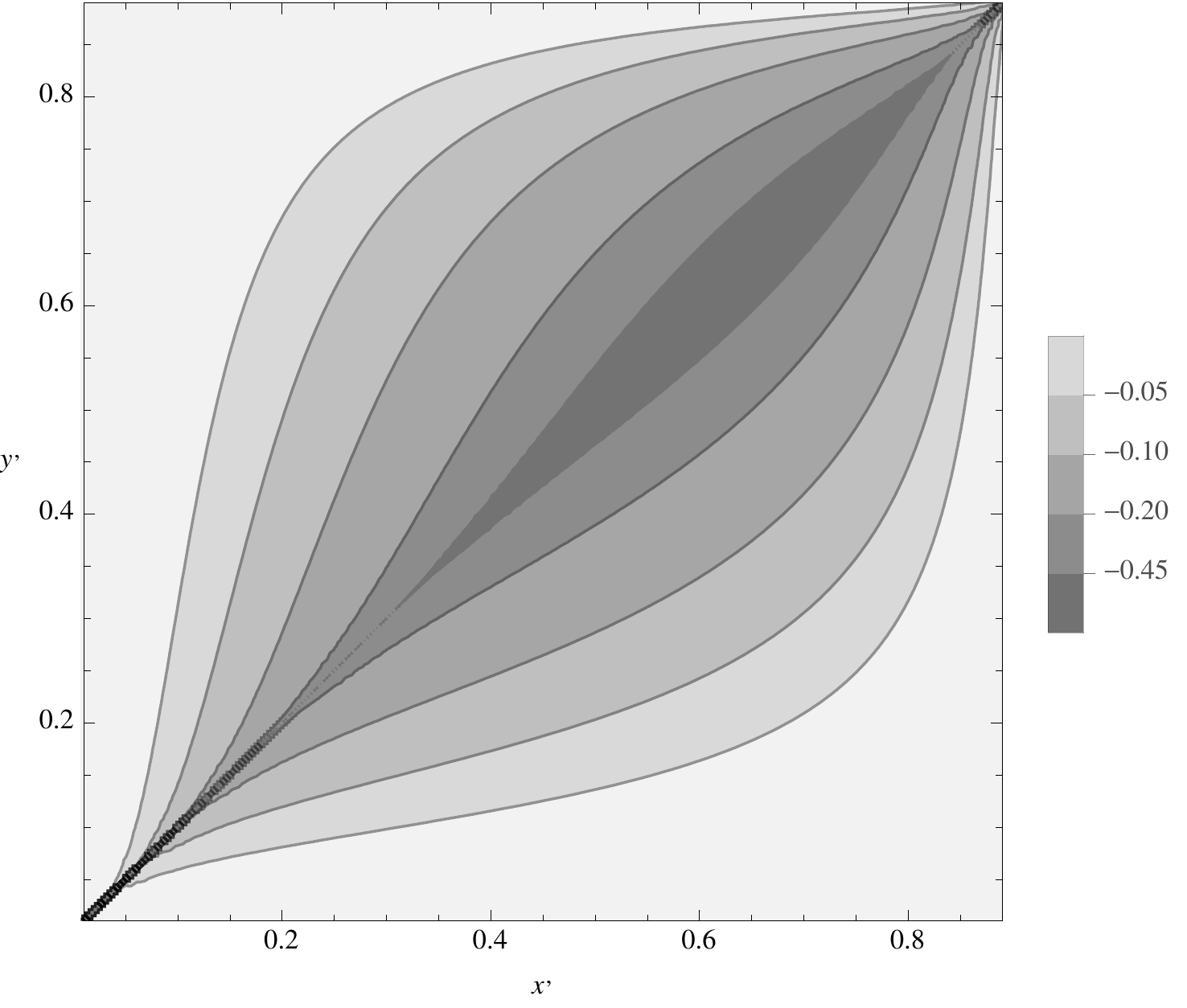} 
\captionsetup{width=0.9\textwidth}
\caption{A contour plot of the kernel $N(x',y')$ giving the non local part of the modular Hamiltonian for two intervals for $\eta=9/10$. $N(x',y')$ has a logarithmic divergence along the diagonal.}
\label{contor}
\end{center}
\end{figure}

A simplification in the structure of the non local kernel $N(x',y')$ arises if we take into account that the integration $
\int_{A_1'}dx'\, v_s^k(x')=0$. Hence, using (\ref{masas}) we could write (\ref{ene})  as 
\bea
N(x',y')=-\int_0^{x'}d\tilde{x}\, \int_{y'}^\eta d\tilde{y}\, (k_i(\tilde{x},\tilde{y})+k_r(\tilde{x},\tilde{y}))\,,\hspace{.8cm} x'<y'\,, \label{mas}\\
N(x',y')=-\int_{x'}^\eta d\tilde{x}\, \int_0^{y'} d\tilde{y}\, (k_i(\tilde{x},\tilde{y})+k_r(\tilde{x},\tilde{y}))\,,\hspace{.8cm} x'>y'\,.\label{menos}
\eea
In this way we avoid crossing the $\tilde{x}=\tilde{y}$ line in the integration, and therefore the delta functions (\ref{deltauno}) and (\ref{deltados}) do not contribute. Moreover, the integrals are now completely regular and can be done numerically  since we do not have to cross the singular points of the distribution. We checked these expressions coincide with (\ref{ene}).\footnote{We have evaluated (\ref{ene}) extracting the singular contribution and integrating it analytically, and adding the numerical integral of the regular parts.} A contour plot of $N(x',y')$ for $\eta=9/10$ is shown in figure \ref{contor}.

\subsubsection{Summary of the modular Hamiltonian for two intervals}

Summarizing the results, the modular Hamiltonian contains a local part and a non local part
\be
{\cal H}={\cal H}_{\textrm{loc}}+{\cal H}_{\textrm{noloc}}\,.
\ee
The local part, generated by the delta function in (\ref{toje}) and (\ref{tiro}),
 gives a contribution to the modular Hamiltonian that writes on the full region $A$ as
\be
{\cal H}_{\textrm{loc}}=\int_A dx\, \pi\,(\omega'(x))^{-1}\, j(x)^2=\int_A dx\, 2\pi\, (\omega'(x))^{-1}\, T(x)\,,
\ee
where $T(x)=1/2\,j^2(x)$ is the energy density. 
The quantity $\beta(x)=2\pi\, (\omega'(x))^{-1}$ acts as the local inverse temperature multiplying the energy density operator and controls the limit of relative entropy between the vacuum and energetic localized excitations around $x$ \cite{loctemp}. This term, written in terms of the energy density, is equal to the local term of the modular Hamiltonian for the free fermion studied in section 2. This result coincides with general expectations for this term to be universal across two dimensional theories \cite{loctemp,ante}.

The non local part of the modular Hamiltonian is
\bea
{\cal H}_{\textrm{noloc}}&=& \int_{A_1\times A_1} dx\, dy\, j(x)\,N(x,y)\,  j(y)-\int_{A_1\times A_2} dx\, dy\, j(x)\,N(x,\bar{y})\,  j(y)\nonumber\\ &&- \int_{A_2\times A_1} dx\, dy\, j(x)\,N(\bar{x},y)\,  j(y)
 +\int_{A_2\times A_2} dx\, dy\, j(x)\,N(\bar{x},\bar{y})\,  j(y)\,.
\eea
 The relevant kernel, $N(x,y)$, follows from (\ref{mas}), (\ref{menos}), (\ref{kirr}) and (\ref{kreg}). In contrast to the case of the fermion, here it is less singular than the local term. It is given by an integrable function, with at most a $\log|x-y|$ singularity for $x\sim y$. Again in contrast to the fermion case, the modular Hamiltonian is completely non local, the kernel does not vanish identically in any open set of $A\times A$.      

\bigskip

The modular flow is defined as the unitary transformation ${\cal O}(\tau)=e^{i \tau {\cal H}}{\cal O}e^{-i \tau {\cal H}}$
of the operators localized in $A$. For an operator linear in the current ${\cal O}(\tau)=\int dx\, \gamma(x,\tau)\,j(x)$ we have the linear flow equation
\be
\partial_\tau\gamma(x,\tau)=- \, \beta(x)\,  \partial_x\,\gamma(x,\tau)-2 \, \int_A dy\, N(x,y)\, \partial_y\gamma(y,\tau) \,.
\ee
Then, if we start with a $\gamma(x,0)$ localized in an open interval inside the first interval $A_1$ and separated away from the boundary of $A$ by a finite distance, for any $\tau\neq 0$ the function $\gamma(A,\tau)$ will spread everywhere in both intervals $A_1$ and $A_2$. The expectation value $\langle {\cal O}^\dagger(\tau) {\cal O}(\tau)\rangle$ should be finite.  From the correlator (\ref{cotto}) it follows that the Fourier transform $\hat{\gamma}(p,\tau)$ of $\gamma(x,\tau)$ should satisfy $\int dp\, |\hat{\gamma}(p,\tau)|^2 |p|<\infty$. This does not allow for a sharp discontinuity in the test function $\gamma(x,\tau)$ at the boundary of $A$, that would give $\hat{\gamma}(p,\tau)\sim p^{-1}$. However, 
 a term falling like the boundary behaviour of $N(x,y)$ (that is, falling to zero with $x$ as $x\rightarrow 0$ faster than any power of $\log(x)^{-1}$) can keep the test function in the space of allowed functions.
     
Of course, the eigenvectors of the modular Hamiltonian kernel diagonalize the modular flow.  If we decompose the test function 
\bea
\gamma(x,\tau) &=& \sum_{k=1}^2\,\int ds\,  \tilde{\gamma}_k(s,\tau)\,  u^k_s(x) \,,\\
\tilde{\gamma}_k(s,\tau) &=& \int_A dx\,  v_s^k(x)^*\, \gamma(x,\tau)\,,
\eea
the flow equation gets diagonalized according to (\ref{ene})
\be
\tilde{\gamma}(s,\tau)= e^{2 \pi \,s\, i}\,\tilde{\gamma}(s,0) \,.
\ee

\section{Failure of Haag duality for two intervals}
\label{failure}

For simplicity, in this section we  compactify the line and think in a circle $S^1$ instead. Let us consider two disjoint intervals $A_1$, $A_2$ in the circle, and the complement is formed by two disjoint intervals, that we call $A_3$, $A_4$.  We call $A'$ to the complement of a region $A$, hence $(A_1\cup A_2)'=A_3\cup A_4$. If the global state is pure one usually assumes
\be
S(A_1\cup A_2)=S(A_3\cup A_4)\,.\label{syy}
\ee
This is equivalent to
\be
I(A_1,A_2)=I(A_3,A_4)+(S(A_1)+S(A_2)-S(A_3)-S(A_4))\,.
\ee
Taking into account the mutual information is a function of the cross ratio and that the entropies for single intervals are equal to the one of complementary intervals, we can express this relation for the model in the line as
\be
I(\eta)=I(1-\eta)+ \frac{1}{6} \log\left(\frac{\eta}{1-\eta}\right)\,,\label{cucas}
\ee
or, equivalently \cite{cteo}
\be
U(\eta)=U(1-\eta)\,.
\ee
That is, the symmetry property for the entropy of complementary regions (\ref{syy}) gives the symmetry of the function $U(\eta)$. This symmetry has also been shown as a consequence of modular invariance of two dimensional CFT's. We have seen this symmetry is not present for the free chiral current. This is not a problem of the continuum limit, the same happens in a finite lattice as we will show in the next section. In the rest of this section we explain how it is possible.  

The essential reason is that we have two basic choices for algebras of the two intervals. Let us call ${\cal A}'$ to the commutant of the algebra ${\cal A}$, that is, the set of all bounded operators that commute with all operators of ${\cal A}$. If we take a set of operators $\cal S$, the smallest algebra containing this set is ${\cal S}''$. This is the generated algebra by ${\cal S}$. The first choice of algebra for two intervals is just the algebra generated by $j(x)$ for $x$ in the region, and we call it,
\be
{\cal A}^{(1)}(A_1\cup A_2)=\left\{\left(\int dx \, \alpha(x)\, j(x)\right)\,,\, \alpha(x)=0\,\,\, \textrm{for}\,\, x\in (A_1\cup A_2)' \right\}''\,.   \label{sd}
\ee
Here we are smearing the field inside the region.\footnote{There is an additional technical point in getting algebras of bounded operators, that can be thought as doing the spectral decomposition of the smeared fields and taking the algebra of the projectors, or, equivalently, taking the algebra of the exponentials of these operators. We are assuming this step when writing the generated algebra as in (\ref{sd}).}

There is another algebra that can be attached to the two intervals, which consist of adding another operator to the generating set of the previous algebra. This is given by
\be
\Phi_{12}= \int dx\,f(x)\, j(x)\,,  \label{dose}
\ee
where $f(x)$ is any smooth function such that
\be
f(x) = \left\{ \begin{array}{cc} 1\hspace{1cm} x\in A_3\,\\
0 \hspace{1cm} x\in A_4 
\end{array}\right. \,.
\ee
The second algebra is generated by ${\cal A}^{(1)}(A_1\cup A_2)$ and this new operator\footnote{More correctly one should add the spectral projectors of this operator or the unitaries $e^{i a \Phi_{12}}$ for all $a\in R$.}
\be
{\cal A}^{(2)}(A_1\cup A_2)= {\cal A}^{(1)}(A_1\cup A_2) \vee \{\Phi_{12}\}\,.
\ee
It is evident that any two operators $\Phi_{12}$ given by two different functions $f(x)$ with the above  properties will differ by an element of ${\cal A}^{(1)}(A_1\cup A_2)$, and therefore the choice of $f(x)$ satisfying these properties will not change the algebra ${\cal A}^{(2)}(A_1\cup A_2)$.
Note we can define ${\cal A}^{(2)}(A_1\cup A_2)$ as the algebra of $A_1\cup A_2$ because $\Phi_{12}$ actually commutes with all $j(x)$ for $x \in A_3\cup A_4$. This is precisely  because $f(x)$ is constant in $A_3\cup A_4$, and  the commutation relations (\ref{commut}). Notice these are two possible different algebra choices for the same underlying theory and for the same region.

We call the mode $\Phi_{12}$  a ``long link" joining $A_1$ with $A_2$, since being the integral of $\partial_+ \phi$, it is equivalent to some difference of the field $\phi$ localized inside the two intervals. However, in this model the field $\phi$ does not actually exist, and the only way to write this operator is through an integral of the current as in (\ref{dose}).

There is still the possibility of adding a long link crossing the interval $A_4$, rather than $A_3$ as in $\Phi_{12}$. However, a sum of these two long links is (modulo operators in ${\cal A}^{(1)}(A_1\cup A_2)$) equivalent to the integral of the current on the circle $\oint dx\,j(x)$. This last operator commutes with all the algebra and generates the global center. In a specific Hilbert space representation this operator will take a fixed value and then would be equivalent to a multiple of the identity operator. Hence the two options for the long link are actually equivalent.

Now, we can define in analogous way the algebras ${\cal A}^{(1)}(A_3\cup A_4)$ and ${\cal A}^{(2)}(A_3\cup A_4)$. A long link crossing $A_1$ for example would be
\be  
\Phi_{34}=\int dx\, g(x)\, j(x)\,,
\ee
with
\be
g(x) = \left\{ \begin{array}{cc} 1\hspace{1cm} x\in A_1\,\\
0 \hspace{1cm} x\in A_2 
\end{array}\right. \,.
\ee
It follows from the commutation relations that
\be
\left[\Phi_{12},\Phi_{34} \right]=i \int dx\, f(x) g'(x) =i \neq 0\,.
\ee
Therefore ${\cal A}^{(2)}(A_1\cup A_2)$ and ${\cal A}^{(2)}(A_3\cup A_4)$ do not commute, and cannot be the algebras of complementary regions. Instead we have\footnote{This can be shown easily in the finite lattice model of the next section.} 
\bea
\left({\cal A}^{(1)}(A_1\cup A_2)\right)' &=& {\cal A}^{(2)}(A_3\cup A_4)\,,\\
\left({\cal A}^{(2)}(A_1\cup A_2)\right)' &=& {\cal A}^{(1)}(A_3\cup A_4)\,.
\eea

In general, the algebras of complementary regions $V$ and $\bar{V}$ commute, i.e. ${\cal A}(\bar{V})\subseteq ({\cal A}(V))' $. If this inclusion is instead an equality, i.e. ${\cal A}(\bar{V})= ({\cal A}(V))'$, it is said that the model satisfies  Haag duality for the region $V$. 
Hence, in order to have Haag duality for a two interval region in this model one should choose the long link for one of the pairs of intervals and not for the complementary one. This prescription  necessarily  does not treat in an equivalent way all pair of intervals. Another perhaps more disturbing consequence of this choice is that the algebra   ${\cal A}^{(2)}$ containing the long link is not additive, meaning that it is not the generated algebra by the algebras of the two intervals. This is because the long link does not belong to the algebra generated by the single intervals. Hence we can have additivity at the expense of Haag duality, or viceversa, but not both properties together. The natural choice is the 
${\cal A}^{(1)}(A_1\cup A_2)$ because it can be consistently assigned to any two interval regions, and because of the additivity property, it is the only choice that allows the definition of mutual information. 

The relation (\ref{syy}) holds for example when  the complementary regions correspond to tensor product of full algebras in a lattice and the global state is pure. This situation will always give Haag duality. In the present case, failure of Haag duality indicates one can still think in terms of tensor products (in a regularized model) but where one of the factors does not have an interpretation in terms of the algebra of two intervals; it contains additional ``long-link'' operators. Hence, for the chiral current eq. (\ref{syy}) fails not because of the global state is not pure but because of failure of Haag duality, i.e., the commutant of the algebra of a region is not the algebra of the complementary region.     

It is worth to notice that the algebra of the current $j(x)$ is a subalgebra of the free massless chiral fermion. It is precisely the subalgebra generated by the fermion current $\psi^\dagger \psi$. The fermion is an extension of the current algebra, and it is one that satisfies Haag duality for two intervals.\footnote{The fields with odd fermionic number are included in the regions by  slightly generalizing the causality requirement \cite{longosuper}. The fermion model is also additive.} This is why for the fermion $U(\eta)=U(1-\eta)$ since trivially $U(\eta)\equiv 0$.  It is by reducing the theory to the current algebra that we run into this particular trouble. The current subalgebras of the free fermion for two intervals are of the form ${\cal A}^{(1)}$, since the long link (\ref{dose}) measures the  charge in $A_3$ and does not commute with the charged fermion field.    

This failure of Haag duality for two intervals in general chiral conformal models has been associated to an algebraic index ($\mu$-index) of the  inclusion of subalgebras ${\cal A}^{(1)}(A_3\cup A_3)\subset \left({\cal A}^{(1)}(A_1\cup A_2)\right)'={\cal A}^{(2)}(A_3\cup A_4)$  \cite{kawa}. This index should also determine the amount of asymmetry in the mutual information \cite{longo} as\footnote{The paper \cite{longo} proves this relation for subalgebras of the free fermion field and conjectures its greater validity for chiral CFT models.}
\be
U(0)-U(1)=\frac{1}{2}\log(\mu)\,.
\ee
In the present model we have seen this is divergent, in accordance with the fact that the $\mu$-index of the current is infinity\footnote{We thank Roberto Longo for clarifications on the material of this section.}.

\subsection{Two chiralities and restoration of Haag duality}

From the point of view of CFT in  $d=2$ it might seem strange that we could have Haag duality violation for two intervals and hence $U(\eta)\neq U(1-\eta)$. This property can be derived from modular invariance of the twist operators giving place to the Renyi entropies \cite{moduinvar}. The reason is that Haag duality can be restored by adequately combining two chiral theories. 

Let us look at the example of the massless limit of a free massive scalar. The usual local algebra for a massive scalar in $d=2$ is  Haag dual and additive for two ``diamonds'', corresponding to two intervals in the spatial line at $t=0$. At zero mass the zero mode of the scalar field has divergent mean quadratic variation and has to be removed. Then what remains are spatial and time derivatives of the field. With these we can form $\partial_\pm \phi$, the two chiral currents. For a single diamond, the algebra is then equivalent to the one of two decoupled chiral currents in an interval. For two diamonds however, the algebra also contains the difference $\phi(x_2)-\phi(x_1)$, with $x_1$ and $x_2$ belonging to the two different diamonds. We can take $x_1,x_2$ on the two intervals at $t=0$. This is 
\be
\phi(x_2)-\phi(x_1)=\int_{x_1}^{x_2} dx\, \partial_x \phi(0,x)= \int_{x^+_1}^{x^+_2}dx^+\, \partial_+ \phi(x)- \int_{x^-_1}^{x^-_2}dx^-\, \partial_- \phi(x)\,.\label{sdf}
\ee
 Hence, the two diamonds contain the difference of long link operators corresponding to the two chiral algebras. However, they do not contain the sum of these long link operators, and therefore the chiralities do not decouple for two diamonds. This is the reason these algebras for the two diamonds are compatible with the ones of the two complementary diamonds: the chiral long link operators do not commute to each other but their sum does, since commutators come out with opposite sign. Thinking in terms of the field differences (\ref{sdf}) this commutation is evident. Therefore these algebras have the same form for all pairs of diamonds and Haag duality is retained.\footnote{Note that the tensor product of the two chiral scalars in two intervals gives a different algebra, which can be associated with four (non space-like separated) diamonds in Minkowski space.} 
 However, without the zero mode, additivity is lost in this example. The massless limit of the mutual information of two intervals is divergent as $I\sim 1/2 \log(-\log(m))$ \cite{review}.

\section{The chiral current in the lattice} \label{numerico}
For doing numerical simulations we put the model in a lattice. We take the lattice Hamiltonian
\be
H=\frac{1}{2}\sum f_i^2\,,
\ee
and the commutator
\be
[f_i,f_j]=i (\delta_{j,i+1}-\delta_{j,i-1})\equiv i \, C\,.\label{c}
\ee

Let us take a periodic system, and in order for $C$ to be invertible, we take an even number $N=2 n$ of points. 
The eigenvectors of the commutator are given by phase factors
\be
\sum_j C_{j l} e^{i k l}= 2 i \sin(k) e^{i k j}\,,\hspace{1cm}k=\frac{2 \pi m}{N}\,,\,\,\, m=-(n-1),\cdots,n\,.  
\ee
From here it follows that, defining the following variables for $0<k<\pi$ (that is, $m\in (1,n-1)$),  
\bea
\phi_k &=& \frac{1}{\sqrt{N \sin(k)}} \sum_j \cos(k j)\, f_j \,,\\
\pi_k &=& \frac{1}{\sqrt{N \sin(k)}} \sum_j \sin(k j)\,f_j \,,
\eea
they are canonical conjugates, $[\phi_k,\pi_{k'}]=i \,\delta_{k,k'}$. There are another two variables that form a global center of the algebra since they commute with all other elements,
\be
\psi_0=\frac{1}{\sqrt{N}}\sum_j f_j\,, \hspace{1cm} \psi_\pi= \frac{1}{\sqrt{N}}\sum_j (-1)^j f_j\,,
\ee
The inverse relation is
\be
f_j=\frac{2}{\sqrt{N}}\sum_{0<k<\pi} \sqrt{\sin(k)}\left( \cos(k j) \, \phi_k + \sin(k j) \pi_k \right) +\frac{\psi_0}{\sqrt{N}}+\frac{\psi_\pi (-1)^j}{\sqrt{N}}\,.
\ee
The Hamiltonian then writes in these new variables
\be
H=\sum_{0<k<\pi} \sin(k) (\phi_k^2+\pi_k^2)+\frac{1}{2} \psi_0^2+\frac{1}{2} \psi_\pi^2\,.\label{hmy}
\ee
This gives for the vacuum state $\langle \phi_k^2\rangle=\langle \pi_k^2\rangle=1/2$, $\langle \phi_k \pi_k\rangle=i/2$. The center can take any value and we set $\psi_0=\psi_\pi=0$. Hence we impose these relations as a constraint. In this way we get a pure vacuum state and a global algebra without center.
 The full system has now $n-1$ degrees of freedom: $n-1$ coordinates and $n-1$ momentum variables.  

We see from (\ref{hmy}) that we have two sets of low energy degrees of freedom, for $k\sim 0$ and $k\sim \pi$. Hence the system shows doubling of degree of freedom in the continuum, analogous to the usual fermion doubling. This is also the reason we have two commuting operators $\psi_0$ and $\psi_\pi$.  

The correlator of the original variables $F(i-j)=\langle f_i f_j\rangle$ is given by
\bea
F(x) &=& \frac{1}{N}\frac{\cos^2\left(\frac{\pi x}{2}\right)\sin\left(\frac{2 \pi}{N}\right)}{\sin\left(\frac{\pi (x+1)}{N}\right)\cos\left(\frac{\pi(x-1+N/2)}{N}\right)} \,,\hspace{.7cm} |x|\neq 1\,,\\
F(x) &=& \frac{i}{2} C (x)\,, \hspace{3.9cm} |x|=1\,.
\eea
In the limit of a large circle $N\rightarrow \infty$ we have 
\bea
F(x) &=& -\frac{1+(-1)^x}{\pi (x^2-1)}\,,\hspace{.7cm} |x|\neq 1\,, \label{f}\\
F(x) &=& \frac{i}{2} C (x) \,, \hspace{3.9cm} |x|=1\,.
\eea

The entropy of a region follows from (\ref{ona}), (\ref{one}). 
We first check numerically the entropy for an interval. We calculate the matrices (\ref{c}) and (\ref{f}) for intervals of length $R=10 k$ with $k=1,...,20$. We fit the pairs $(R_k,S(R_k))$ with  $c_0+c_{\log}\log k+c_{-1}\frac{1}{k}+c_{-2}\frac{1}{k^2}$
obtaining the logarithmic coefficient $c_{\log}=1/3$ with high precision. Notice this coefficient is twice the expected one for the chiral current model. This reflects the doubling on the lattice. 

To calculate the mutual information between two intervals of length $a$ and $b$ separated by a distance $c$, that is $I(A,B)=S(A)+S(B)-S(A\cup B)$, we need the entropies of the single intervals $S(A)$ and $S(B)$ and the entropy of the two intervals $S(A\cup B)$. Each of these entropies are calculated using (\ref{ona}, \ref{one}). In the continuum limit the mutual information is a function of the cross ratio $\eta$,  $I(A,B)=I(\eta)$ where $\eta$ is defined as
\be
\eta=\frac{a.b}{(a+c)(b+c)}\,,
\ee
in accordance to (\ref{crossratio}).
For a given cross ratio, we repeat the calculation for different configurations that differ one from another just by a dilatation with parameter $k=2,4,...,20$. We then fit the pairs $(k\,,I_k(\eta))$ with  $c_0+c_{-1}\frac{1}{k}+c_{-2}\frac{1}{k^2}+c_{-3}\frac{1}{k^3}$
and take the constant coefficient $c_0$ as the continuum limit of the mutual information for the lattice model, which is twice the chiral current model due to doubling. We then take $I(\eta)=c_0/2$. We repeat the same procedure for different values of $\eta$ obtaining the red points showed in figure \ref{figufi}.

In doing simulations for this model it is important that, if $N$ is finite, we take the total number of points even $N=2n$, and, in order not to have a center, the subsystems need to have an even number of points or variables (intervals of even size). This is because half of them are coordinates and half are momentum. The complementary subsystems automatically must have equal entropy because the global state is pure. For example, for an interval of size $2k$ in a circle of size $2n$, the commutant is an interval of size $2n-2k-2$, because there are two points in the complementary region adjacent to the interval that do not commute with the original interval. The entropies are indeed equal. 
When we consider two interval regions the commutant algebra contains a long link as explained in the previous section. In the lattice it will contain two long links operators, because of the doubling.  
More precisely,  these commutant algebras with long links for two intervals are of the form: all points in the closed intervals $[a_1,b_1]\cup [a_2,b_2]$, and two long links, given by the sums $\sum f_i$ and $\sum (-1)^i f_i$, where the sums are over all the points in the open interval $(b_1,a_2)=(b_1+1,\cdots,a_2-1)$. The long links crossing the other gap between the intervals are related to these by elements of the algebra of the intervals and the global constraints, and hence they do not give additional contributions. The counting of degrees of freedom is as follows: for a circle of $2n$ points, if the original intervals have $2k_1$  and $2k_2$ points, the commutant will have $2n-2k_1-2k_2-4$ points plus two long links. This gives  a total of $2n-2k_1-2k_2-2$ linearly independent operators. This is precisely (twice) the complementary number of degrees of freedom: $2n-2$ is twice the total number of degrees of freedom in the lattice. 

We have checked the entropies of complementary algebras of two intervals are equal in the circle. The entropy for the two intervals with the long links $\tilde{S}$ can also be completed to form a kind of ``mutual information'', eliminating UV divergences in the continuum, as
\be
\tilde{I}(A_1,A_2)=S(A_1)+S(A_2)-\tilde{S}(A_1\cup A_2)\,.
\ee
The equality of the entropies for two intervals and the one of the complementary region including the long links,  
\be
S(A_1\cup A_2)=\tilde{S}(A_3\cup A_4)\,,
\ee
can be completed with single interval entropies to form a relation between the mutual informations 
\be
\tilde{I}(\eta)=I(1-\eta)+\frac{1}{6}\log\left(\frac{\eta}{1-\eta}\right)\,.\label{illvsi}
\ee
We can define the $U$ function for the entropies with the long link
\be
\tilde{I}(\eta)=-\frac{1}{6}\log(1-\eta)+\tilde{U}(\eta)\,.
\ee
Then relation (\ref{illvsi}) is just the complementary relation for the $U(\eta)$ 
\be
\tilde{U}(1-\eta)=U(\eta)\,.
\ee
These two should be symmetric and equal for a model with Haag duality for two intervals but this is not the case in the present model.
We have also checked numerically the relation (\ref{illvsi}) in the infinite lattice. 
For that we calculate $\tilde{I}(\eta)$ and we note that convergence to the continuum limit is much improved for this case using  the fitting function as $c_0+c_{-1/2}\frac{1}{k^{1/2}}+c_{-3/2}\frac{1}{k^{3/2}}$, instead of integer powers, as we increase the global size $k$ of the region. The continuum limit again correspond to the coefficient $\tilde{I}(\eta)=c_0/2$. 

\section{Final Remarks}
\label{conclusions}

We have diagonalized the vacuum density matrix for a chiral scalar in two intervals. The modular Hamiltonian contains the usual local term given by an integral of the energy density times a position dependent inverse temperature. This term is identical to the free fermion one, and very probably is universal for all theories in $d=2$. In addition, there is a non local term. This is given by a quadratic expression in the current with a locally integrable kernel which does not vanish in any open set of $A\times A$. Hence the modular Hamiltonian is completely non local in contrast to the fermion case. 

The mutual information does not have the symmetry property (\ref{cucas}), and the origin of this is the failure of duality for two intervals.  

We treated the case of two intervals. More intervals could in principle be treated in a similar fashion, but the expressions will depend on a higher number of cross rations, and besides the Hypergeometric and Appell functions that parametrize the eigenvectors should be replaced by higher dimensional Lauricella functions. 

It would be interesting to understand why the fermion modular Hamiltoninan is quasilocal while the one of the current is completely non local. The technical reason is that one eigenvector of the bosonic model ($u_2$) has a dependence of the eigenvalue $s$ that is not simply a phase factor $e^{i s \omega(x)}$. For the fermion field and any number of intervals, or the current field in the single interval case, this same phase factor determines completely the dependence of all eigenvectors in $s$. In this case the modular flow has a simple geometrical picture as a translation in the variable $\omega$, $\omega\rightarrow \omega+ 2 \pi \tau$ \cite{modu}.  
Perhaps a reason for the fermion to be special is the multilocal symmetries described by Rehren and Tedesco \cite{rehren2}.  

We have shown that the current mutual information is smaller than the fermion one because the former model is a subalgebra of the later. It would be interesting to explore other consequences of this inclusion. For example, the difference of modular Hamiltonians ${\cal H}_\psi-{\cal H}_j$ between these models should be a positive operator. We can compute the expectation value of this difference of operators in a state generated from the vacuum by acting with a unitary in $A$, for example a coherent state $e^{i \int dx\, \gamma(x) j(x)}|0\rangle$. The local contribution vanish in the difference of expectation values of the two modular Hamiltonians and we get an inequality involving exclusively the non local parts of ${\cal H}_\psi$ and ${\cal H}_j$. 

\section*{Acknowledgments}
We thank discussions with Karl-Henning Rehren, Erik Tonni, Gabriel Wong, and specially with Stefan Hollands and Roberto Longo. We also thank Erik Tonni for pointing out several typos in the previous version of the paper. We thank the Galileo Galilei Institute for Theoretical Physics for the hospitality and the INFN for partial support during the completion of this work.
This work was partially supported by CONICET, CNEA and Universidad Nacional de Cuyo, Argentina. H.C. acknowledges support from an It From Qubit grant of the Simons Foundation.

\appendix

\section{Formulas for Gaussian state}
\label{modularscalar}
Formulas for the entropy and modular Hamiltonian for Gaussian states in the algebra of canonical commutation relations in terms of coordinate and momentum correlators are described for example in \cite{review}. We need here a slightly more general approach where we describe the algebra by $2n$ hermitian operators $f_i$, $i=1\cdots 2n$, with a general non degenerate numerical commutator 
\be
\left[f_i,f_j\right]= i C_{ij}\,,
\ee
where $C$ is antisymmetric and real. 
This general case treated for example in \cite{Erik1,Erik2,sorkin}. 
The state is Gaussian with hermitian correlator
\be
F_{ij}=\langle f_i f_j\rangle\,. 
\ee
and we have
\be
C_{ij}=2 \, \textrm{Im}(F_{ij})\,.
\ee
With an orthogonal matrix $O$ we can write $C$ in the form
\be
O \, C\, O^{T}= \left(
\begin{array}{cc}
0 & D\\
-D & 0
\end{array}
\right)\,,
\ee 
where $D$ is a diagonal $n \times n$ matrix with positive elements. 
Another transformation allow us to write this matrix into the canonical form
\be
 M=Q \,O \, C\, O^{T}\, Q= \,\left(
\begin{array}{cc}
0 & 1\\
-1 & 0
\end{array}
\right)\,,
\ee 
where 
\be
Q=\left(
\begin{array}{cc}
D^{-1/2} & 0\\
0 & D^{-1/2}
\end{array}
\right)\,.
\ee
Accordingly  
\be
\vec{\Phi}=Q \, O \, \vec{f}=\left(\phi_1,
\hdots ,
\phi_n ,
\pi_1 ,
\cdots ,
\pi_n
 \right)^{T}
\ee
is a vector of field and momentum with canonical commutation relations. We write
\be
Q \,O \, F\, O^{T}\, Q=\,\left(
\begin{array}{cc}
X & i/2\\
-i/2 & P
\end{array}
\right)\,,\label{assu}
\ee
where $X$ and $P$ are the matrices of correlators of the field and momentum respectively. In writing (\ref{assu}) we are assuming there is no real part of the off diagonal blocks, and this is consequence of time-inversion invariance of the state, that is an anti-unitary symmetry mapping $\phi_i\rightarrow \phi_i$, $\pi_i\rightarrow -\pi_i$.
 In terms of the original variables $f_i$  it is an anti-unitary symmetry mapping linearly $f_i\rightarrow T_{ij} f_j$, with $T$ real and  $T C T^{T}=-C$.  

We can choose the system of eigenvectors of $XP$ and $PX$,
\bea
XP u_\nu &=& \nu^2 \, u_\nu\,\\
PX v_\nu &=& \nu^2 \, v_\nu\,,
\eea
such that
\bea
X u_\nu &=& \nu \, v_\nu\,,\\
P v_\nu &=& \nu\, u_\nu\,.
\eea
They can be normalized with
\be
\langle u_\nu| v_{\nu'}\rangle=\delta_{\nu,\nu'} \,.
\ee
We have $\nu\ge 1/2$ because of the uncertainty relations.

The matrix  
\be
V=-i C^{-1}\, F-\frac{1}{2}=
O^{T} Q \left(
\begin{array}{cc}
0 & i P\\
-i X & 0
\end{array}
\right) Q^{-1} O\,,\label{ona}
\ee 
has eigenvalues $\pm |\nu|$ corresponding to the eigenvectors $O^T Q (\mp i u_\nu,v_\nu)^{T}$. Hence the formula of the entropy in terms of the $X,P$ correlators \cite{review}
\be
S=\textrm{tr}\left((\sqrt{XP}+1/2)\log(\sqrt{XP}+1/2)-(\sqrt{XP}-1/2)\log(\sqrt{XP}-1/2)\right)\,, 
\ee
writes
\be
S=\textrm{tr}\, (V+1/2) \log|V+1/2|=\textrm{tr}\,  \Theta(V)\,\left((V+1/2)\log(V+1/2)+(1/2-V)\log(V-1/2)\right)\,. \label{one}
\ee
This was first shown in \cite{sorkin}. Analogously, the Renyi entropies defined by
\be
S_n=\frac{1}{1-n}\,\log(\textrm{tr}\,\rho^n)
\ee
are given by
\be
S_n=\frac{1}{n-1}\textrm{tr}\,\Theta(V)\, \log \left[(V+1/2)^n-(V-1/2)^n\right]\,.\label{renyi4}
\ee
The modular Hamiltonian writes \cite{review}
\be
{\cal H}=\vec{\Phi}^{T} 
\left(\begin{array}{cc}
g(PX) P & 0\\
0 &  g(XP) X
\end{array}\right)
 \vec{\Phi}\,,
\ee
where 
\be
g(y)= \frac{1}{2\sqrt{y}}\log\left(\frac{\sqrt{y}+1/2}{\sqrt{y}-1/2}\right)\,.
\ee
In the present notation we have
\be
{\cal H}= -i \vec{f}^T g(V^2)\,V\,   C^{-1}   \vec{f}\,.
\ee

\end{document}